%% file: iclr2027.tex
\documentclass{article}
\usepackage{iclr2027_conference,times}

\input{math_commands.tex}

\usepackage{hyperref}
\usepackage{url}
\usepackage{booktabs}
\usepackage{amsmath}
\usepackage{amsthm}
\usepackage{amssymb}
\usepackage{bm}
\usepackage{graphicx}
\usepackage{subcaption}
\usepackage{algorithm}
\usepackage{algorithmicx,algcompatible}
\usepackage{float}
\usepackage{multirow}
\usepackage{enumitem}
\usepackage[table]{xcolor}
\usepackage{microtype}

\usepackage{array, algpseudocode}

\newtheorem{theorem}{Theorem}
\newtheorem{lemma}[theorem]{Lemma}

\newtheorem{proposition}[theorem]{Proposition}
\theoremstyle{definition}

\theoremstyle{remark}

\title{Asymmetric Speculative Decoding: Unlocking In-Context Learning in Draft Models}

\author{
  Guang Huang \\
  HKUST(GZ) \\
  \texttt{ghuang817@connect.hkust-gz.edu.cn}
  \And
  Zeyi Wen \\
  HKUST(GZ), HKUST \\
  \texttt{wenzeyi@hkust-gz.edu.cn}
}
\iclrfinalcopy
\begin{document}

\maketitle
\lhead{Preprint}

\input{sections/abstract}
\input{sections/introduction}

\input{sections/related_work}
\input{sections/method}
\input{sections/experiments}

\input{sections/conclusion}

\subsection*{AI Use Statement}

We used generative AI tools for text revision, experimental code assistance, and visualization code generation, but all core contributions, experimental design, and final manuscript were produced by the authors, who take full responsibility for the work.

\subsection*{Ethics Statement}

This work presents a training-free inference optimization technique that does not involve human subjects or introduce ethical concerns beyond those of existing LLM deployment, and provides no additional capability for harmful content generation beyond what the underlying models already possess.

\subsection*{Reproducibility Statement}

All implementation details, model configurations, and experimental settings are provided in the paper, and we will release complete source code upon publication to ensure full reproducibility with publicly available models and datasets.

\bibliography{references}
\bibliographystyle{iclr2027_conference}

\appendix
\input{sections/appendix}

\end{document}

%% file: math_commands.tex
\usepackage{amsmath,amsfonts,bm}

\def\eqref#1{equation~\ref{#1}}

\def\1{\bm{1}}

\DeclareMathAlphabet{\mathsfit}{\encodingdefault}{\sfdefault}{m}{sl}
\SetMathAlphabet{\mathsfit}{bold}{\encodingdefault}{\sfdefault}{bx}{n}



%% file: sections/abstract.tex
\begin{abstract}
Speculative decoding addresses Large Language Model (LLM) inference latency by first drafting and then verifying in parallel. 
Yet practical speedups fall far short of theoretical bounds, especially on reasoning tasks. 
We identify the root cause as \textit{task isomorphism}: existing methods provide identical prompts to both models, forcing the draft model to solve problems beyond its capacity. 
This produces high rejection rates that eliminate parallelization gains. 
We discover that acceptance rates increase monotonically during generation, revealing a latent in-context learning (ICL) effect wherein the draft model progressively aligns with the target distribution by learning from verified tokens. 
Controlled experiments with varying context window sizes confirm this improvement stems from ICL-driven alignment rather than token difficulty.
Motivated by this finding, we propose \textbf{Asymmetric Speculative Decoding (\textsc{AsymSpec})}, a training-free framework that decouples the prompts provided to draft and target models. 
By supplying the draft model with task-relevant example solutions, we reformulate its objective from reasoning to simulation, amplifying the ICL effect from the first token. 
We prove AsymSpec produces exact samples and converges exponentially to the target distribution. 
Across reasoning, code generation, and summarization tasks, AsymSpec achieves 2.29× to 4.95× speedup over autoregressive decoding, outperforming standard speculative decoding without model retraining.
\end{abstract}

%% file: sections/introduction.tex
\section{Introduction}

The autoregressive nature of Large Language Model (LLM)~\citep{vaswani2017attention,brown2020language,touvron2023llama} necessitates sequential token generation, resulting in high inference latency that limits practical deployment~\citep{pope2023efficiently,kwon2023efficient}. Speculative decoding~\citep{leviathan2023fast,chen2023accelerating} addresses this bottleneck by employing a small draft model to propose candidates and a larger target model verifies in parallel.
Although this approach promises substantial speedups, actual performance often falls short of theoretical limits, particularly in reasoning-intensive tasks.~\citep{xia2024unlocking,sun2024spectr,cai2024medusa,li2024eagle}.

We identify the root cause as \textit{task isomorphism}: existing frameworks provide both models with identical prompts, requiring them to solve the same problem independently. 
When presented with a complex reasoning task~\citep{wei2022chain,cobbe2021training}, the draft model's output deviates from the target distribution due to its limited capacity.
High rejection rates eliminate the parallelization advantage~\citep{sun2024spectr,kim2024speculative}.

\begin{figure}[t]
    \centering
    \begin{subfigure}[b]{0.32\textwidth}
        \centering
        \includegraphics[width=\textwidth]{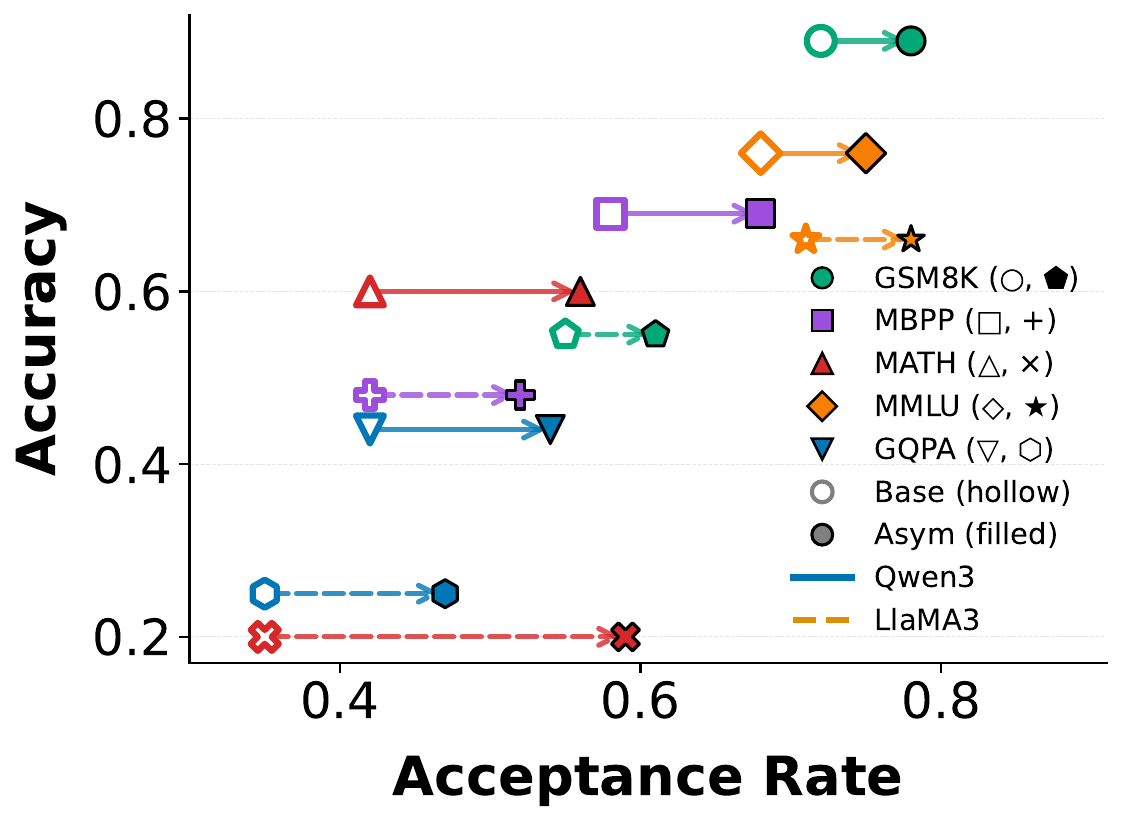}
        \caption{\textbf{Task isomorphism bottleneck}}
        \label{fig:task_isomorphism}
    \end{subfigure}
    \hfill
    \begin{subfigure}[b]{0.32\textwidth}
        \centering
        \includegraphics[width=\textwidth]{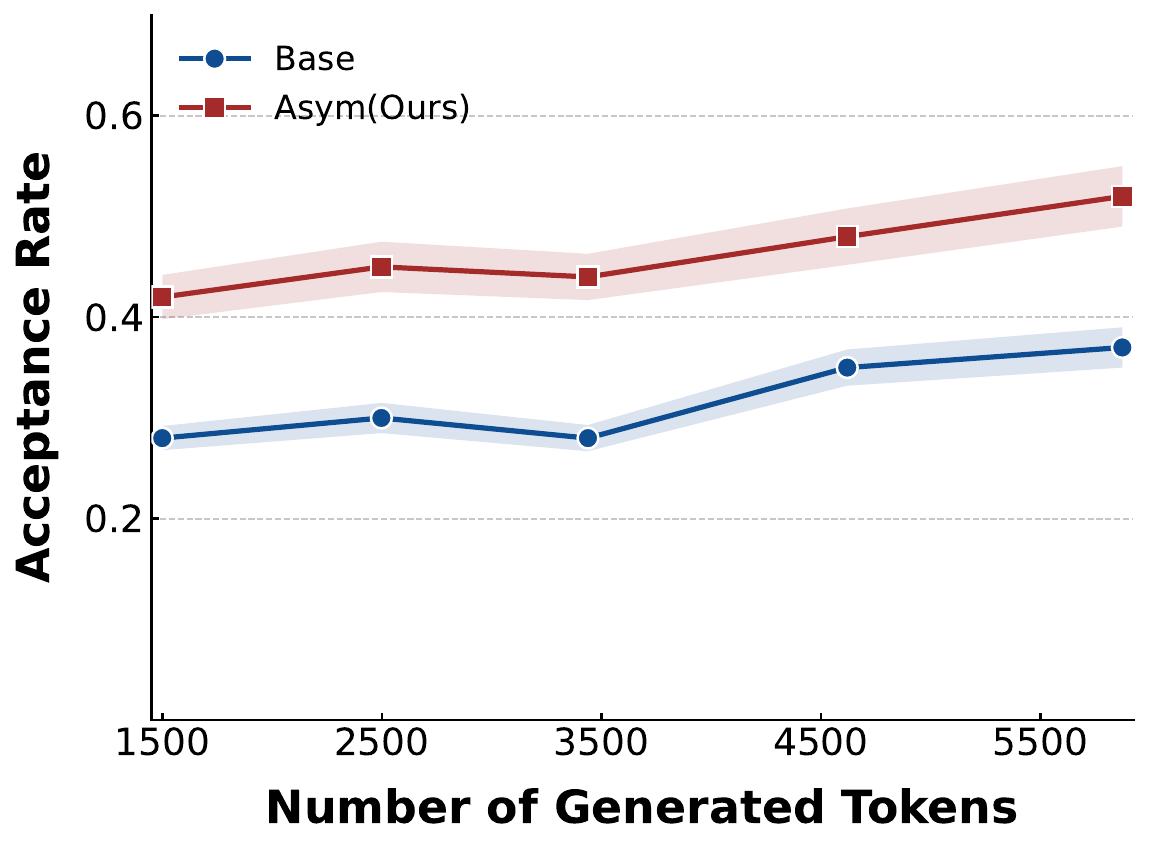}
        \caption{\textbf{Acceptance rates improvement}}
        \label{fig:acceptance_evolution}
    \end{subfigure}
    \hfill
    \begin{subfigure}[b]{0.32\textwidth}
        \centering
        \includegraphics[width=\textwidth]{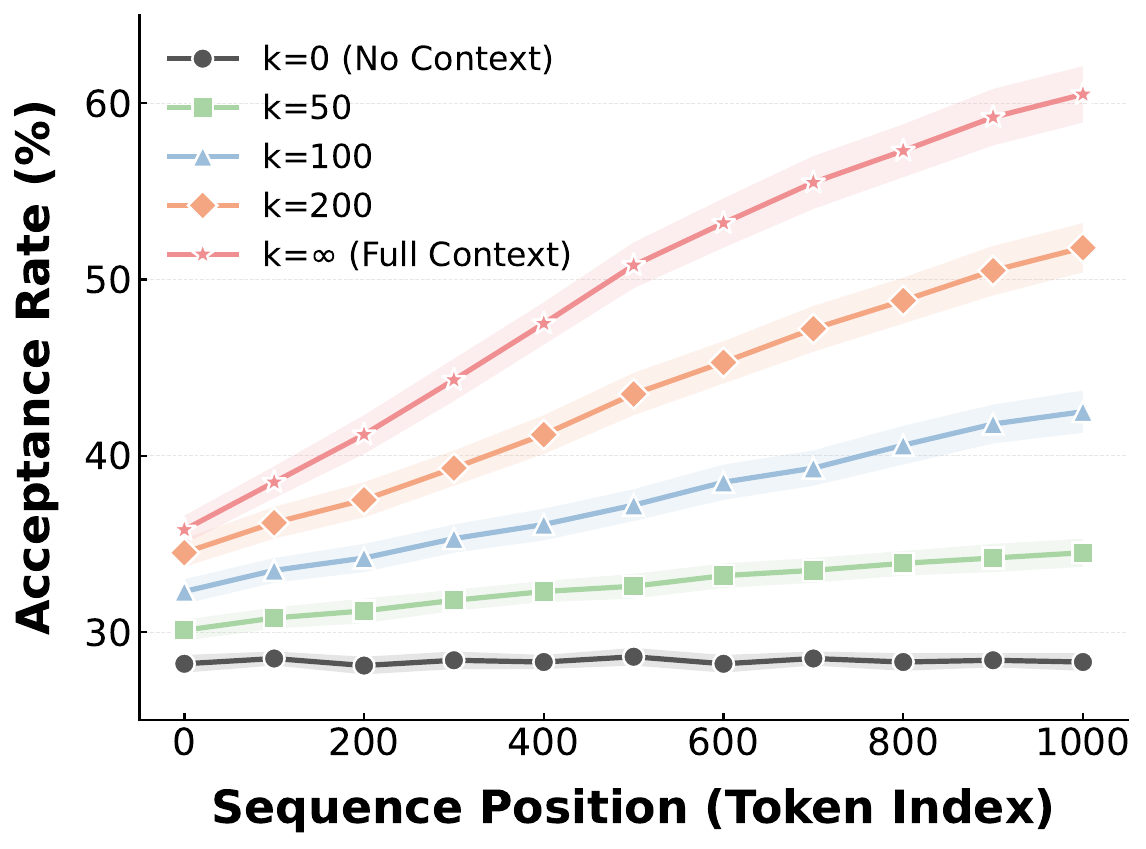}
        \caption{\textbf{Controlling for token difficulty}}
        \label{fig:icl_confirmation}
    \end{subfigure}

    \caption{
    \textbf{Empirical motivation for asymmetric speculative decoding.}
    (a) Acceptance rates under Base (standard speculative decoding) and Asym (Ours) across five benchmark tests and two model series. Each point represents a task-model combination.
    (b) Acceptance rates as a function of generated token in standard speculative decoding (Base) and \textsc{AsymSpec} (Ours).
    (c) Acceptance rates as a function of sequence position under different context window sizes.
    }
    \label{fig:motivation}
\end{figure}

To validate this hypothesis, empirical analysis across reasoning benchmarks and model families reveals that the task isomorphism bottleneck manifests as substantial variation in acceptance rates under symmetric prompting (Figure~\ref{fig:task_isomorphism}). Critically, even when draft models demonstrate reasonable standalone task accuracy, acceptance rates remain suboptimal. This gap indicates that the bottleneck lies in distributional alignment rather than reasoning capability. However, asymmetric prompting with behavioral exemplars consistently improves acceptance across diverse tasks, suggesting a mechanism orthogonal to problem-solving capacity. This raises a fundamental question: why does providing exemplars enable distributional alignment?

Investigation of acceptance dynamics throughout the generation process provides critical insight (Figure~\ref{fig:acceptance_evolution}). In standard speculative decoding, acceptance rates are not uniformly poor but instead improve progressively as generation proceeds. This pattern suggests the draft model learns from accumulated verified history $\mathbf{y}_{<t}$, transitioning from independent problem-solving to trajectory simulation. We interpret this as a latent \textit{in-context learning (ICL) effect}~\citep{brown2020language,dong2022survey,min2022rethinking}. Specifically, the draft model aligns with the target distribution through implicit demonstrations~\citep{dai2022why,xie2021explanation} provided by verified tokens. 

Controlled experiments confirm that this improvement stems from ICL-driven distributional convergence rather than decreasing token difficulty (Figure~\ref{fig:icl_confirmation}). Acceptance rates under varying context window sizes reveal that larger contexts produce steeper improvements across sequence positions. Larger contexts provide more verified history for ICL. With minimal context, acceptance rates remain relatively flat. As context expands, the rate of improvement accelerates markedly. This confirms that ICL-driven alignment, enabled by accumulated verified tokens, underlies the observed phenomenon rather than intrinsic token predictability.

These findings suggest a natural solution: if the draft model can learn to align through accumulated examples, why wait for this alignment to emerge gradually? We propose \textbf{Asymmetric Speculative Decoding (\textsc{AsymSpec})}, a framework that provides the ICL effect upfront by decoupling the prompts given to draft and target models. The target model receives the original reasoning task, while the draft model receives an asymmetric prompt enriched with task-relevant example solutions. The framework maintains separate KV caches for each model and employs asymmetric synchronization during verification to preserve exact target distributions.


We establish theoretical correctness by proving that \textsc{AsymSpec} produces exact samples through rejection sampling (Theorem~\ref{thm:correctness}) and that draft-target distributions converge exponentially as verified history grows (Proposition~\ref{prop:convergence}). Across reasoning, code generation, summarization, and instruction following tasks, \textsc{AsymSpec} achieves 2.29$\times$ to 4.95$\times$ speedup over autoregressive decoding on Qwen3 and LLaMA families, consistently outperforming standard speculative decoding and training-free baselines. Compared to training-based method, \textsc{AsymSpec} achieves competitive throughput without requiring model retraining. The method demonstrates robustness across various exemplar pools, sampling temperatures and batch sizes. Ablation studies reveal an optimal exemplar lengths that balance ICL effectiveness against prefill overhead.

In summary, our contributions are threefold:
\begin{itemize}[leftmargin=*, parsep=0pt, itemsep=2pt, topsep=2pt]
    \item We identify task isomorphism as the fundamental limitation of speculative decoding on reasoning tasks and reveal a latent in-context learning effect: draft models progressively align with target distributions through accumulated verified context, but this alignment emerges too slowly to benefit early-stage generation.
    \item We propose \textsc{AsymSpec}, a training-free framework that activates the ICL effect from the initial token by providing asymmetric prompts with behavioral exemplars, reformulating the draft model's objective from independent reasoning to trajectory simulation.
    \item We prove that \textsc{AsymSpec} maintains exact correctness through rejection sampling and achieves exponential distributional convergence. Experiments across diverse benchmarks demonstrate 2.29$\times$ to 4.95$\times$ speedups, matching training-based methods without parameter updates.
\end{itemize}

%% file: sections/related_work.tex
\section{Related Work}

\textbf{Speculative Decoding.} Speculative decoding~\citep{leviathan2023fast, chen2023accelerating} accelerates LLM inference by using a compact draft model to predict tokens verified in parallel by the target model. Recent advances pursue higher acceptance through architectural enhancements (specialized draft models~\citep{kim2024speculative}, multi-head expansions~\citep{cai2024medusa, li2024eagle, eagle3}, knowledge distillation~\citep{hinton2015distilling}) or draft-free mechanisms (self-speculation~\citep{zhang2023drafting}, lookahead~\citep{fu2024break}, cascaded verification~\citep{ssd}). All existing methods operate under \textit{symmetric prompting} with identical inputs. We identify task isomorphism as the bottleneck and propose asymmetric task reformulation to unlock draft model ICL capacity.

\textbf{In-Context Learning.} ICL enables LLMs to adapt to novel tasks via exemplars without parameter updates~\citep{brown2020language}, with theoretical work showing ICL implicitly performs gradient descent~\citep{dai2022why} or Bayesian inference~\citep{xie2021explanation}. While extensively applied to improve standalone model accuracy~\citep{wei2022chain, zhou2022least, chen2021evaluating}, ICL's potential in \textit{collaborative inference} remains unexplored. We leverage ICL to transform draft models from independent solvers into adaptive simulators that predict target model behavior.

\textbf{Asymmetric Multi-Model Systems.} Asymmetric designs have been explored for task-level routing in Mixture-of-Experts~\citep{fedus2022switch} and confidence-based cascades~\citep{schuster2022confident}, but focus on \textit{task-level specialization} rather than \textit{token-level collaboration}. Our framework is the first to systematically exploit prompt asymmetry for speculative decoding, achieving efficiency gains without architectural modifications or training. More related work details are in appendix \ref{app:related_work}.

%% file: sections/method.tex
\section{Methodology}
\label{sec:method}

We formally introduce \textbf{Asymmetric Speculative Decoding (\textsc{AsymSpec})}. We begin by formulating the standard speculative decoding paradigm to highlight its inherent limitations (\S\ref{subsec:prelim}). Next, we present empirical evidence revealing a latent in-context learning effect (\S\ref{subsec:motivation}), which motivates our framework. We then detail the structural decoupling mechanism (\S\ref{subsec:framework}), establish its theoretical correctness through rigorous analysis (\S\ref{subsec:theory}), and analyze computational complexity (\S\ref{subsec:complexity}).

\subsection{Preliminaries and Problem Formulation}
\label{subsec:prelim}

\textbf{Standard Speculative Decoding.} Let $\mathcal{M}_T$ denote a large target model and $\mathcal{M}_D$ a smaller draft model. Given query $\mathbf{x}$, standard speculative decoding~\citep{leviathan2023fast} operates in two phases. First, $\mathcal{M}_D$ generates a candidate sequence $\tilde{Y} = [\tilde{y}_1, \dots, \tilde{y}_\gamma]$ autoregressively. Second, $\mathcal{M}_T$ evaluates all candidates in parallel via rejection sampling~\citep{leviathan2023fast,chen2023accelerating}  , accepting token $\tilde{y}_i$ with probability:
\begin{equation}
    \alpha = \min \left( 1, \frac{P_T(\tilde{y}_i | \mathbf{x}, \mathbf{y}_{<t}, \tilde{Y}_{<i})}{P_D(\tilde{y}_i | \mathbf{x}, \mathbf{y}_{<t}, \tilde{Y}_{<i})} \right)
\end{equation}
Speedup depends critically on acceptance rates. High acceptance allows parallel verification to amortize the target model's cost, while low acceptance degrades to sequential generation.

\subsection{Motivation: Uncovering the In-Context Learning Bottleneck}
\label{subsec:motivation}

Standard speculative decoding enforces a task isomorphism constraint: the draft model must independently solve the same reasoning problem as the target model. We demonstrate that this constraint creates a fundamental bottleneck and reveal a latent in-context learning mechanism that motivates our asymmetric design.

\textbf{The Task Isomorphism Bottleneck.} Acceptance rates across five reasoning benchmarks reveal substantial variation under symmetric prompting (30\% to 70\%) despite reasonable standalone accuracy (Figure~\ref{fig:task_isomorphism}). Critically, even when the draft model achieves 60\% accuracy on MATH, acceptance rates remain at 42\%. This gap indicates the bottleneck lies in distributional alignment rather than reasoning capability. The draft model reaches correct answers through divergent trajectories that the target model rejects. Asymmetric prompting with behavioral exemplars consistently improves acceptance across all tasks.

\textbf{Progressive Alignment Through Generation.} Acceptance rates rise monotonically from 20\% at early tokens to 40\% by sequence end under standard speculative decoding (Figure~\ref{fig:acceptance_evolution}), suggesting the draft model progressively learns from verified history $\mathbf{y}_{<t}$ through in-context learning (ICL). Experiments with varying context window sizes confirm this: larger windows produce steeper acceptance increases, while minimal context yields flat rates (Figure~\ref{fig:icl_confirmation}). This disentangles ICL-driven alignment from token difficulty effects.

\textbf{Design Implication.} Standard speculative decoding incurs a cold-start penalty while waiting for ICL to emerge organically. Our solution provides behavioral exemplars upfront, activating the ICL effect from the initial token.

\begin{figure}[t]
    \centering
    \includegraphics[width=0.9\linewidth]{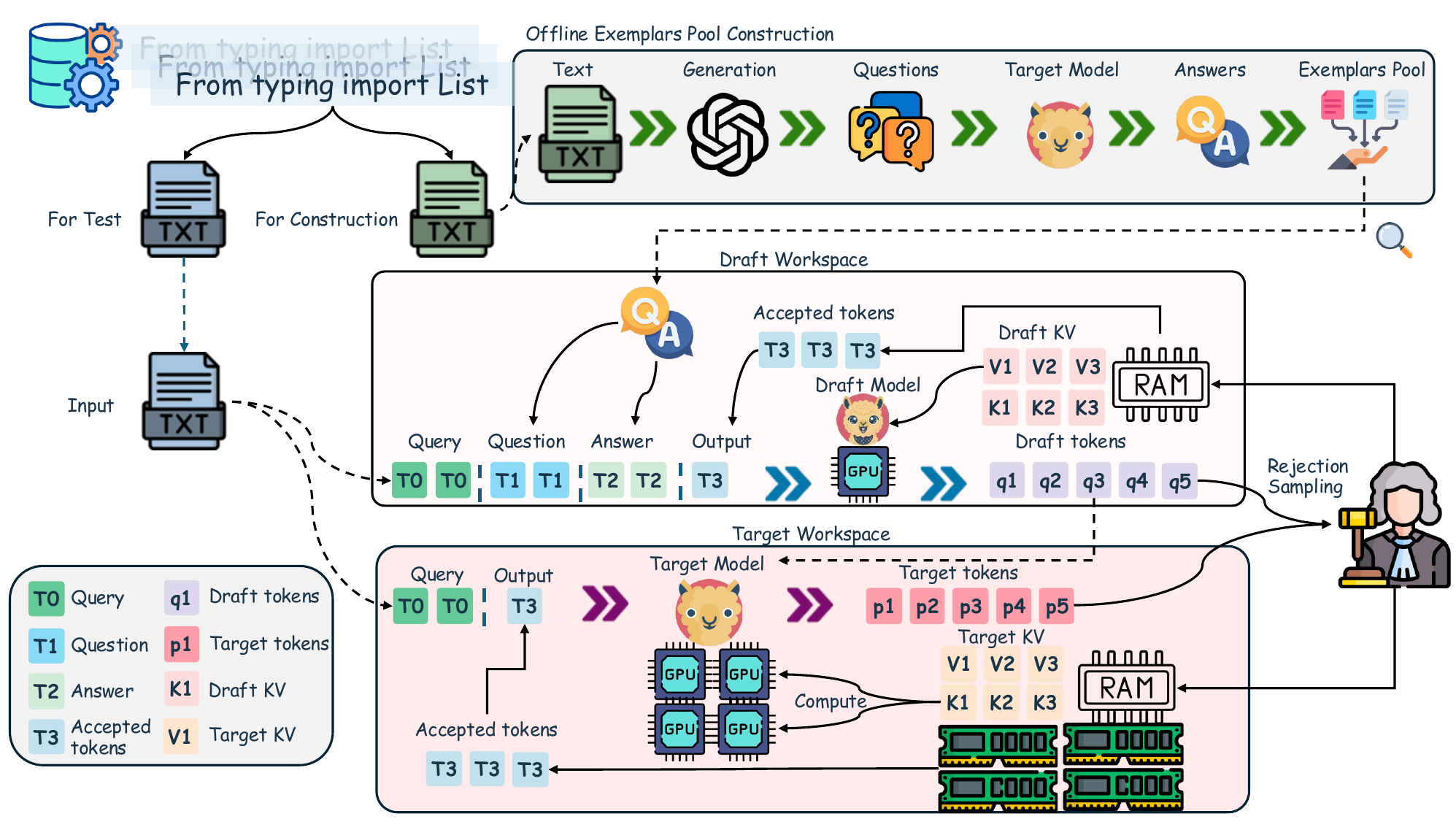}
    \caption{The overall framework of AsymSpec.}
    \label{fig:framework}
\end{figure}

\subsection{The \textsc{AsymSpec} Framework}
\label{subsec:framework}

Based on the insight that $\mathcal{M}_D$ excels at trajectory simulation rather than independent reasoning, we introduce \textsc{AsymSpec}, which abandons the task isomorphism constraint by structurally decoupling the contextual states of draft and target models. The overall framework is illustrated in Figure~\ref{fig:framework}.

\textbf{Offline Exemplars Pool Construction.} 
Since there are no open-source datasets containing demonstrations of the target model's output behavior, we construct an exemplars pool $\mathcal{E}_{\text{pool}}$ offline. We use the first 50\% of each dataset for pool construction and the remaining 50\% for evaluation, with no overlap. Inspired by prompt caching in production APIs, we sample task-relevant questions, generate responses using $\mathcal{M}_T$, and store $(c_i, r_i)$ pairs as behavioral templates. During inference, we retrieve $K$ exemplars from $\mathcal{E}_{\text{pool}}$ based on task alignment to construct $\mathcal{E}_{\text{sim}}$. This one-time construction enables prompt caching when exemplars are reused across queries.

\textbf{Asymmetric Prompt Formulation.}
Instead of forcing $\mathcal{M}_D$ to comprehend the complex logic of $\mathbf{x}$, we alter its generation objective through an asymmetric prompt:
\begin{equation}
    \mathcal{P}_{\text{draft}} = \left[ \mathcal{I}_{\text{role}} \oplus \mathcal{E}_{\text{sim}} \oplus \mathbf{x} \right]
\end{equation}
where $\oplus$ denotes concatenation. This comprises: (1) $\mathcal{I}_{\text{role}}$, a system instruction shifting the draft model's mode from solving to format imitation; (2) $\mathcal{E}_{\text{sim}} = \{(c_1, r_1), \dots, (c_K, r_K)\}$, $K$ task-aligned behavioral exemplars activating the ICL effect from the first token; and (3) $\mathbf{x}$, the original query. This explicitly conditions the draft distribution $P_D^{\text{sim}}(\cdot \mid \mathcal{P}_{\text{draft}}, \mathbf{y}_{<i})$ on trajectory simulation rather than task resolution.

\textbf{Decoupled Prefill and Drafting.}
Because $\mathcal{P}_{\text{draft}} \neq \mathbf{x}$, the framework maintains two physically distinct Key-Value cache prefixes to preserve exact target distributions. During prefill, KV tensors are computed independently:
\begin{align}
    \mathbf{KV}_{T}^{(0)} &= \text{Prefill}_T(\mathbf{x}) \\
    \mathbf{KV}_{D}^{(0)} &= \text{Prefill}_D(\mathcal{P}_{\text{draft}})
\end{align}
In the drafting phase, $\mathcal{M}_D$ auto-regressively generates $\gamma$ candidates:
\begin{equation}
    \tilde{y}_i \sim P_D^{\text{sim}}(\tilde{y}_i \mid \mathbf{KV}_{D}^{(i-1)}), \quad \mathbf{KV}_{D}^{(i)} = \mathbf{KV}_{D}^{(i-1)} \cup \text{KV}_D(\tilde{y}_i)
\end{equation}
Let $\tilde{Y} = [\tilde{y}_1, \dots, \tilde{y}_\gamma]$ denote the draft sequence.

\textbf{Parallel Verification and Asymmetric Synchronization.}
The target model $\mathcal{M}_T$ processes candidates $\tilde{Y}$ in a single parallel forward pass, retrieving from its unmodified task cache $\mathbf{KV}_{T}$:
\begin{equation}
    P_T(\tilde{y}_i \mid \mathbf{x}, \mathbf{y}_{<t}, \tilde{Y}_{<i}) = \text{Forward}_T \left( \tilde{Y} \mid \mathbf{KV}_{T}^{(t-1)} \right)
\end{equation}
We employ standard rejection sampling: a draft token $\tilde{y}_i$ is accepted if $r_i \leq \min \big( 1, \frac{P_T(\tilde{y}_i \mid \mathbf{x}, \mathbf{y}_{<t}, \tilde{Y}_{<i})}{P_D^{\text{sim}}(\tilde{y}_i \mid \mathcal{P}_{\text{draft}}, \mathbf{y}_{<t}, \tilde{Y}_{<i})} \big)$, where $r_i \sim \mathcal{U}(0, 1)$.

When rejection occurs at the $k$-th token, the framework accepts prefix $\tilde{Y}_{<k}$ and $\mathcal{M}_T$ samples a corrective token $y_{\text{corr}}$ from its residual distribution. The target cache advances by absorbing the verified sequence:
\begin{equation}
    \mathbf{KV}_{T}^{(t+k)} = \mathbf{KV}_{T}^{(t-1)} \cup \text{KV}_T(\tilde{Y}_{<k} \oplus y_{\text{corr}})
\end{equation}
The draft cache performs asymmetric rollback, truncating the invalid trajectory $\tilde{Y}_{k:\gamma}$ and synchronizing with $y_{\text{corr}}$:
\begin{equation}
    \mathbf{KV}_{D}^{(t+k)} = \text{Truncate}(\mathbf{KV}_{D}^{(t+\gamma)}, k-1) \cup \text{KV}_D(y_{\text{corr}})
\end{equation}
This guarantees that both models enter the next iteration with identical historical text sequences.

\subsection{Theoretical Analysis}
\label{subsec:theory}

\textbf{Correctness.} Despite using different prompts, \textsc{AsymSpec} produces exact samples from the target distribution.

\begin{theorem}[Correctness of \textup{\textsc{AsymSpec}}]
\label{thm:correctness}
For any verified history $\mathbf{y}_{<t}$, \textup{\textsc{AsymSpec}} produces exact samples from the target distribution:
\begin{equation}
P_{\text{AsymSpec}}(y_t \mid \mathbf{x}, \mathbf{y}_{<t}) = P_T(y_t \mid \mathbf{x}, \mathbf{y}_{<t})
\end{equation}
\end{theorem}

\begin{proof}
The correctness follows from rejection sampling~\cite{chen2023accelerating,leviathan2023fast}. We accept draft token $\tilde{y}_i$ with probability:
\begin{equation}
\alpha = \min \left( 1, \frac{P_T(\tilde{y}_i \mid \mathbf{x}, \mathbf{y}_{<t}, \tilde{Y}_{<i})}{P_D^{\text{sim}}(\tilde{y}_i \mid \mathcal{P}_{\text{draft}}, \mathbf{y}_{<t}, \tilde{Y}_{<i})} \right)
\end{equation}
By the fundamental theorem of rejection sampling~\citep{kim2000monte}, the output distribution equals the target distribution exactly. While the draft model uses asymmetric prompt $\mathcal{P}_{\text{draft}}$ for generation, the target model always evaluates candidates using the original prompt $\mathbf{x}$, maintaining exact correctness.
\end{proof}

\textbf{Effectiveness.} Higher acceptance rates arise from distribution convergence as verified history $\mathbf{y}_{<t}$ grows. Two Transformer properties drive this:

\begin{enumerate}[label=(\roman*), leftmargin=*, parsep=0pt, itemsep=2pt]
    \item \textbf{Attention Decay:} Attention on initial prompt tokens decays exponentially~\citep{press2021train,sun2022length}:
    \begin{equation}
    \sum_{i=1}^{|\mathbf{x}|} \text{Attn}(q_t, k_i) \leq \beta \cdot e^{-\lambda t}
    \end{equation}

    \item \textbf{Lipschitz Continuity:} Output distributions are Lipschitz continuous~\citep{gao2017properties}:
    \begin{equation}
    \text{TV}(P(\cdot \mid h), P(\cdot \mid h')) \leq L \|h - h'\|_2
    \end{equation}
\end{enumerate}



\begin{proposition}[Long-Range Non-Degradation]
\label{prop:convergence}
As verified history length $t$ increases, the discrepancy between the asymmetric and standard draft distributions decays exponentially:
\begin{equation}
\text{TV}\left(P_D(\cdot \mid \mathcal{P}_{\text{draft}}, \mathbf{y}_{<t}), P_D(\cdot \mid \mathbf{x}, \mathbf{y}_{<t})\right) \leq C \cdot e^{-\lambda t}
\end{equation}
\end{proposition}

This explains effectiveness in two stages: early tokens (small $t$) benefit from exemplars $\mathcal{E}_{\text{sim}}$ activating ICL to align with the target model, while later tokens (large $t$) are dominated by verified history $\mathbf{y}_{<t}$ as the asymmetric prefix influence vanishes ($C \cdot e^{-\lambda t} \to 0$) without long-term bias. Please see Appendix~\ref{app:proof} for more details.





%% file: sections/experiments.tex
\section{Experiments}
\label{sec:experiments}

\subsection{Experimental Setup}
\label{subsec:setup}

\textbf{Models.} We evaluate \textsc{AsymSpec} on state-of-the-art open-weight LLMs. Target-draft pairs include Qwen-3 (8B/32B), and LLaMA-3.1 (8B/70B).

\textbf{Datasets.} We select benchmarks spanning mathematical reasoning, code generation, and instruction following: (1) \textit{Mathematics}: AIME~\citep{lightman2023lets}, MATH-500~\citep{hendrycks2021measuring}; (2) \textit{Code}: HumanEval~\citep{chen2021evaluating}; (3) \textit{Summarization}: CNN/DM~\citep{see2017get}; (4) \textit{Instruction}: MT-Bench~\citep{zheng2023judging}. 

\textbf{Baselines.} Our method belongs to the family of training-free methods, which uses a small draft model from the same series. Therefore, we compare our method with Vanilla Autoregressive Decoding and Standard Speculative Decoding (SPS)~\citep{leviathan2023fast,chen2023accelerating}). We also compare it with the state-of-the-art training-free SOTA method, Speculative Decoding (SSD)~\citep{ssd}. We also introduce Eagle-3~\citep{eagle3} as a training-required baseline.

\textbf{Metrics and Implementation.} We measure wall-clock speedup ($S$) relative to vanilla autoregressive decoding and mean acceptance length ($\tau$) per speculative iteration. Experiments were conducted on NVIDIA RTX 4090 and A6000 GPUs. We integrated \textsc{AsymSpec} into \texttt{nano-vllm} and used \textit{PageAttention}~\citep{kwon2023efficient} to manage the key-value cache.

\subsection{Main Results}
\label{subsec:main_results}

\begin{table*}[t]
\centering
\caption{Performance comparison across reasoning and coding benchmarks. Speedup ($S$) and acceptance length ($\tau$) are reported relative to Vanilla AR baseline (1.00$\times$). \textsc{AsymSpec} consistently outperforms Standard Speculative Decoding (SPS) and Speculative-Speculative Decoding (SSD) across all model pairs and tasks, with gains scaling favorably on larger capacity gaps. \textbf{Bold} indicates best performance in each column.}
\label{tab:main_results}
\vspace{2mm}

\resizebox{\textwidth}{!}{
\begin{tabular}{cl cc cc cc cc cc cc}
\toprule
\multirow{2.5}{*}{\textbf{Model}} & \multirow{2.5}{*}{\textbf{Method}} & \multicolumn{2}{c}{\textbf{MATH-500}} & \multicolumn{2}{c}{\textbf{AIME25}} & \multicolumn{2}{c}{\textbf{HumanEval}} & \multicolumn{2}{c}{\textbf{CNN/DM}} & \multicolumn{2}{c}{\textbf{MT-Bench}} & \multicolumn{2}{c}{\textbf{Average}} \\
\cmidrule(lr){3-4} \cmidrule(lr){5-6} \cmidrule(lr){7-8} \cmidrule(lr){9-10} \cmidrule(lr){11-12} \cmidrule(lr){13-14}
& & $S$ & $\tau$ & $S$ & $\tau$ & $S$ & $\tau$ & $S$ & $\tau$ & $S$ & $\tau$ & $S$ & $\tau$ \\
\midrule
\rowcolor{gray!10} & Vanilla & 1.00 & 1.00 & 1.00 & 1.00 & 1.00 & 1.00 & 1.00 & 1.00 & 1.00 & 1.00 & 1.00 & 1.00 \\
\addlinespace[4pt]
\multirow{3}{*}{Q3-8B}
& SPS & 1.81 & 4.87 & 1.75 & 4.75 & 1.80 & 4.82 & 1.67 & 4.68 & 1.71 & 4.58 & 1.74 & 4.74 \\
& SSD & 2.16 & 5.26 & 2.08 & 5.08 & 2.09 & 5.19 & 1.96 & 5.11 & 1.93 & 4.90 & 2.04 & 5.10 \\
& \textbf{Ours} & \textbf{2.45} & \textbf{6.13} & \textbf{2.14} & \textbf{5.40} & \textbf{2.42} & \textbf{6.10} & \textbf{2.41} & \textbf{6.35} & \textbf{2.06} & \textbf{5.38} & \textbf{2.29} & \textbf{5.87} \\

\addlinespace[4pt]
\multirow{3}{*}{Q3-32B}
& SPS & 1.94 & 3.55 & 2.54 & 4.65 & 2.24 & 4.12 & 1.81 & 3.42 & 1.94 & 3.55 & 2.09 & 3.85 \\
& SSD & 2.71 & 3.68 & 4.08 & 5.02 & 4.22 & 5.20 & 2.72 & 3.71 & 2.74 & 3.68 & 3.29 & 4.25 \\
& \textbf{Ours} & \textbf{3.91} & \textbf{5.22} & \textbf{4.32} & \textbf{5.51} & \textbf{4.89} & \textbf{6.11} & \textbf{4.21} & \textbf{5.78} & \textbf{4.03} & \textbf{5.22} & \textbf{4.27} & \textbf{5.57} \\

\midrule
\multirow{3}{*}{L3.1-8B}
& SPS & 1.63 & 5.49 & 1.50 & 5.10 & 1.69 & 5.69 & 1.32 & 4.51 & 1.46 & 4.90 & 1.52 & 5.13 \\
& SSD & 1.88 & 6.03 & 1.70 & 5.55 & 1.96 & 6.26 & 1.87 & 6.21 & 1.60 & 5.30 & 1.80 & 5.87 \\
& \textbf{Ours} & \textbf{2.05} & \textbf{6.63} & \textbf{1.80} & \textbf{5.92} & \textbf{2.05} & \textbf{6.55} & \textbf{2.17} & \textbf{7.02} & \textbf{2.67} & \textbf{6.31} & \textbf{2.14} & \textbf{6.48} \\
\addlinespace[4pt]

\multirow{3}{*}{L3.1-70B}
& SPS & 3.40 & 5.63 & 2.99 & 5.04 & 3.26 & 5.61 & 2.66 & 3.69 & 2.69 & 4.62 & 3.00 & 4.98 \\
& SSD & 5.52 & 6.19 & 4.42 & 5.47 & 4.99 & 6.22 & 2.93 & 4.09 & 3.82 & 4.92 & 4.33 & 5.38 \\
& \textbf{Ours} & \textbf{5.65} & \textbf{6.41} & \textbf{4.53} & \textbf{5.61} & \textbf{5.35} & \textbf{6.68} & \textbf{4.74} & \textbf{6.60} & \textbf{4.52} & \textbf{5.95} & \textbf{4.95} & \textbf{6.25} \\
\bottomrule
\end{tabular}
}
\end{table*}

Table~\ref{tab:main_results} presents speedup and acceptance length across all benchmarks and model pairs. \textsc{AsymSpec} consistently outperforms both Standard Speculative Decoding (SPS) and Speculative-Speculative Decoding (SSD) across all configurations, with performance gains scaling with model capacity. On small model pairs (Qwen-3-8B), \textsc{AsymSpec} achieves 2.29$\times$ average speedup versus 1.74$\times$ for SPS, while on large model pairs (LLaMA-3.1-70B), gains increase to 4.95$\times$ versus 3.00$\times$ for SPS. This scaling behavior reflects the core mechanism: larger capacity gaps amplify the benefits~\citep{hinton2015distilling,gou2021knowledge} of asymmetric prompting, as draft models leverage behavioral exemplars to compensate for reasoning limitations.

\textsc{AsymSpec} substantially improves task consistency compared to symmetric baselines. On Qwen-3-8B, SPS exhibits high variance across tasks (1.67$\times$--1.81$\times$), while \textsc{AsymSpec} narrows this range to 2.06$\times$--2.45$\times$, demonstrating that asymmetric prompting mitigates the task isomorphism trap. Compared to SSD, which achieves a 4.33x speedup on the LLaMA-3.1-70B dataset, \textsc{AsymSpec} delivers a 4.95$\times$ speedup, representing a 0.62$\times$ improvement in speedup ratio.

\subsection{System Robustness}
\label{subsec:robustness}

\begin{figure}[t]
    \centering
    \includegraphics[width=1.0\linewidth]{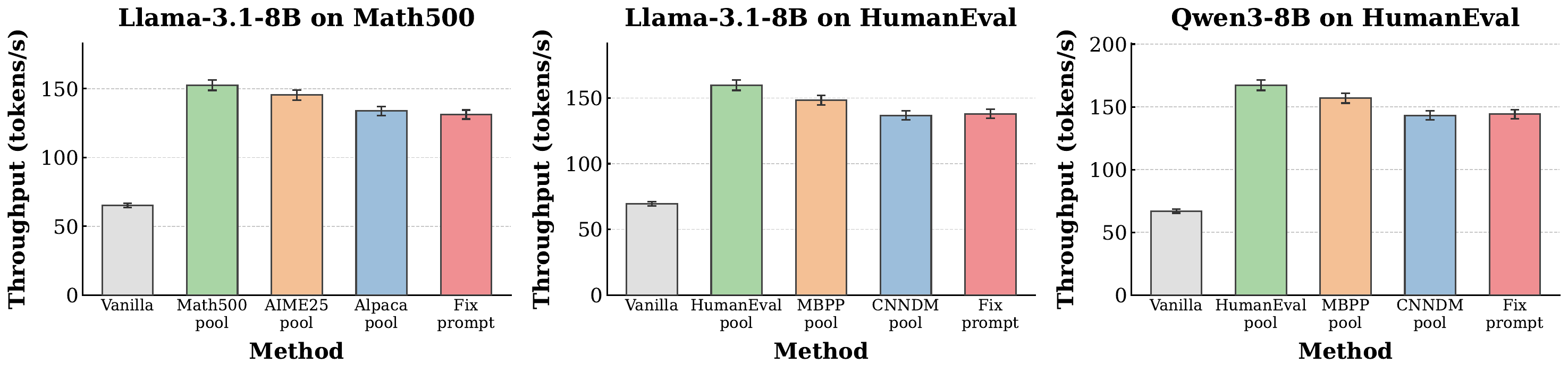}
    \caption{Throughput comparison across different draft pool construction strategies on LLaMA-3.1-8B (MATH-500 and HumanEval) and Qwen-3-8B (HumanEval) at $B=1$, $T=0.0$. Task-specific draft pools consistently achieve the highest throughput, with HumanEval-derived and Math500-derived pools outperforming generic alternatives by 5--8\%.}
    \label{fig:pool_comparison}
\end{figure}

\textbf{Impact of Draft Pool Construction.} 
For HumanEval and MATH-500, we use the first 50\% of each dataset exclusively for exemplar pool construction and reserve the remaining 50\% for evaluation. All pool construction strategies are compared on the same held-out evaluation split.
Figure~\ref{fig:pool_comparison} demonstrates that task-specific draft pools consistently outperform generic alternatives across all configurations on LLaMA-3.1-8B (MATH-500, HumanEval) and Qwen-3-8B (HumanEval). On HumanEval (LLaMA-3.1-8B), the HumanEval-derived pool achieves 159.8 tokens/s versus 148.3 tokens/s for MBPP pool (7.8\% improvement) and 136.7 tokens/s for CNNDM pool. On MATH-500, the Math500-derived pool reaches 152.6 tokens/s compared to 145.4 tokens/s for AIME25 pool (4.9\% improvement) and 133.8 tokens/s for Alpaca pool. Similar patterns hold on Qwen-3-8B HumanEval (167.4 tokens/s vs. 156.9 tokens/s for MBPP, 6.7\% improvement). Notably, even the fixed prompt baseline (137.9--144.2 tokens/s) outperforms vanilla decoding (65.2--69.5 tokens/s) by 2.0--2.2$\times$, confirming that \textsc{AsymSpec}'s core mechanism benefits from any reasonable draft source. The critical advantage lies in capturing domain-specific patterns: code benchmarks benefit from code-derived drafts, while math tasks gain from math-oriented examples, enabling dynamic adaptation to diverse workloads without retraining.

\begin{figure}[htbp]
    \centering
    \includegraphics[width=1.0\linewidth]{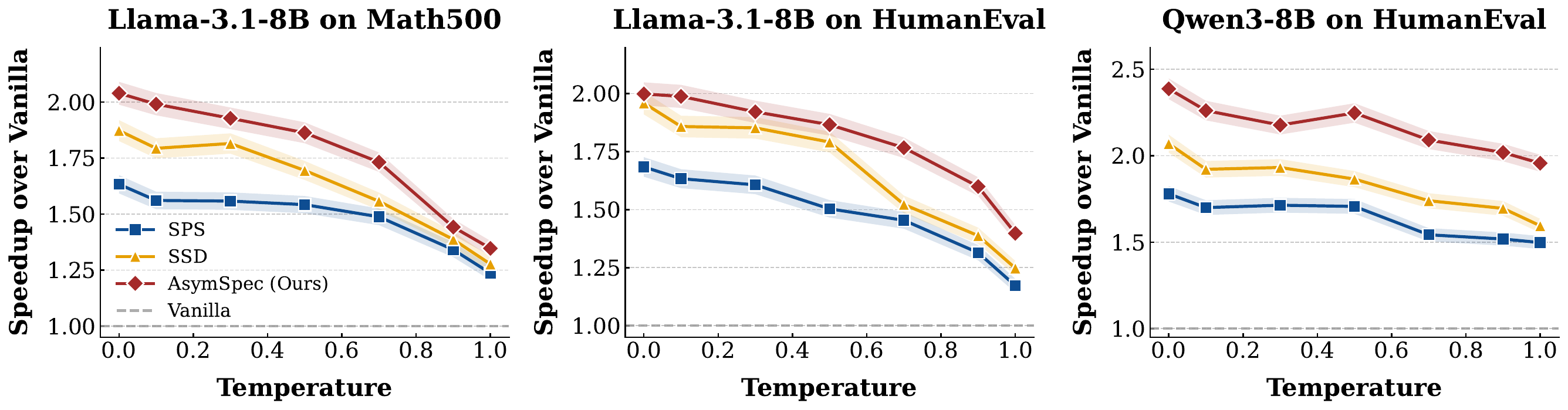}
    \caption{Speedup ratio as a function of sampling temperature. We compare SPS, SSD, and \textsc{AsymSpec} across $T \in [0.0, 1.0]$ on LLaMA-3.1-8B with MATH-500 (left) and HumanEval (middle), and Qwen-3-8B on HumanEval (right). All methods show performance degradation as temperature increases, with \textsc{AsymSpec} maintaining the highest speedup across all settings.}
    \label{fig:temp_robustness}
\end{figure}

\textbf{Temperature Robustness.} Figure~\ref{fig:temp_robustness} demonstrates performance across $T \in [0.0, 1.0]$~\citep{holtzman2020curious} on LLaMA-3.1-8B (MATH-500, HumanEval) and Qwen-3-8B (HumanEval). On MATH-500, Standard Speculative Decoding (SPS) degrades from 1.63$\times$ to 1.23$\times$ (24.2\% drop) and Asynchronous Speculative Decoding (SSD) from 1.87$\times$ to 1.28$\times$ (31.7\% drop), while \textsc{AsymSpec} maintains 2.04$\times$ at $T=0.0$ and 1.35$\times$ at $T=1.0$, preserving substantial advantage over baselines. On HumanEval (LLaMA-3.1-8B), \textsc{AsymSpec} declines from 2.00$\times$ to 1.40$\times$ versus SPS (1.69$\times$ to 1.17$\times$) and SSD (1.96$\times$ to 1.25$\times$). The consistent performance gap validates that behavioral exemplars anchor draft predictions to target patterns, mitigating distribution shift under stochastic sampling.

\begin{figure}[t]
    \centering
    \includegraphics[width=1.0\linewidth]{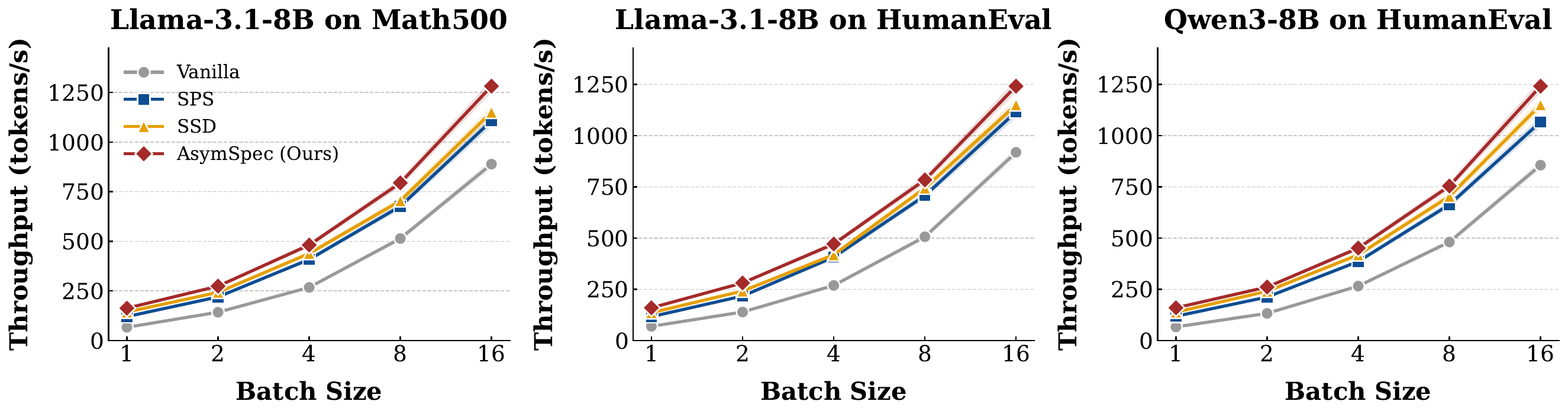}
    \caption{Throughput (tokens/s) as a function of batch size on LLaMA-3.1-8B (MATH-500 and HumanEval) and Qwen-3-8B (HumanEval). AsymSpec consistently outperformed the baseline across all batch sizes, demonstrating its robustness under various deployment constraints.}
    \label{fig:batch_robustness}
\end{figure}

\textbf{Batch Size Scalability.} Figure~\ref{fig:batch_robustness} demonstrates that \textsc{AsymSpec} maintains throughput advantages across batch sizes ($B \in \{1, 2, 4, 8, 16\}$)~\citep{yu2022orca} on LLaMA-3.1-8B (MATH-500, HumanEval) and Qwen-3-8B (HumanEval). On MATH-500, \textsc{AsymSpec} achieves 163.2 tokens/s at $B=1$ versus 120.9 tokens/s for SPS (35.0\% improvement), scaling to 1281.3 tokens/s at $B=16$ versus 1107.2 tokens/s for SPS (15.7\% advantage). Similar trends hold across configurations: on HumanEval (LLaMA-3.1-8B), \textsc{AsymSpec} reaches 1241.3 tokens/s at $B=16$ (11.2\% over SPS). The narrowing relative advantage at larger batch sizes reflects GPU saturation, where the bottleneck shifts from computation to memory bandwidth. However, \textsc{AsymSpec} maintains absolute throughput leadership across all batch sizes, with prefix caching~\citep{zheng2024sglang,kwon2023efficient} amortizing exemplar processing overhead in production deployments. The consistent throughput gains validate that asymmetric prompting scales effectively in production batched inference scenarios. The relative speedup advantage narrows at larger batch sizes (from 35.0\% at $B=1$ to 15.7\% at $B=16$), reflecting GPU saturation where memory bandwidth becomes the bottleneck~\citep{pope2023efficiently,shazeer2019fast}. However, \textsc{AsymSpec} maintains absolute throughput leadership, confirming robustness under varying deployment constraints.

\subsection{Comparison with State-of-the-Art}
\label{subsec:sota}

\begin{figure}[t]
    \centering
    \includegraphics[width=1.0\linewidth]{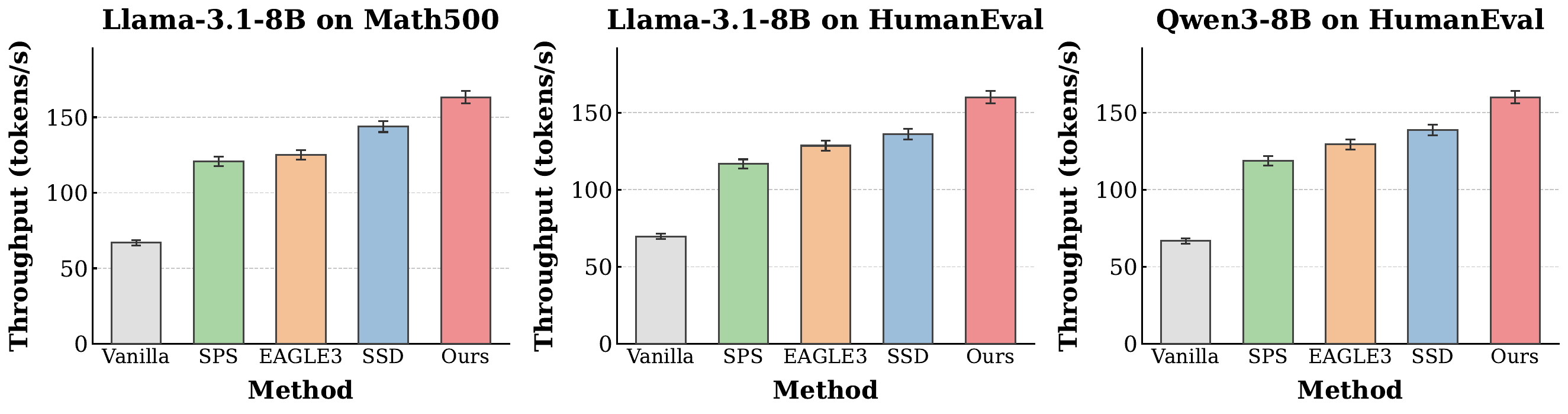}
    \caption{Throughput comparison with state-of-the-art methods on LLaMA-3.1-8B (MATH-500 and HumanEval) and Qwen-3-8B (HumanEval) at $B=1$, $T=0.0$. \textsc{AsymSpec} outperforms training-based EAGLE-3 and asynchronous SSD across all configurations.}
    \label{fig:sota_comparison}
\end{figure}

We compare \textsc{AsymSpec} against training-based EAGLE-3 and asynchronous SSD on identical hardware (NVIDIA A6000, $B=1$, $T=0.0$). Figure~\ref{fig:sota_comparison} shows that \textsc{AsymSpec} achieves competitive or superior throughput across all configurations. On LLaMA-3.1-8B with MATH-500, \textsc{AsymSpec} reaches 163.2 tokens/s versus 125.2 tokens/s for EAGLE-3 (30.3\% improvement) and 143.8 tokens/s for SSD. Similar trends hold on HumanEval: \textsc{AsymSpec} achieves 159.9 tokens/s versus 128.5 tokens/s for EAGLE-3 (24.4\% improvement). These results demonstrate that ICL-based alignment rivals or exceeds specialized draft heads trained on target outputs, while maintaining the advantage of being training-free. The critical distinction lies in deployment costs: training-based methods require substantial upfront investment (EAGLE-3~\citep{eagle3} needs ~100 GPU-hours on 8$\times$ A100) and must retrain for new domains. In contrast, \textsc{AsymSpec} requires only representative samples per task, enabling seamless integration and dynamic adaptation without modifying model weights.

\subsection{Ablation Study: Impact of Extra Prompt Length}
\label{subsec:ablation}

\begin{figure}[t]
    \centering
    \includegraphics[width=1.0\linewidth]{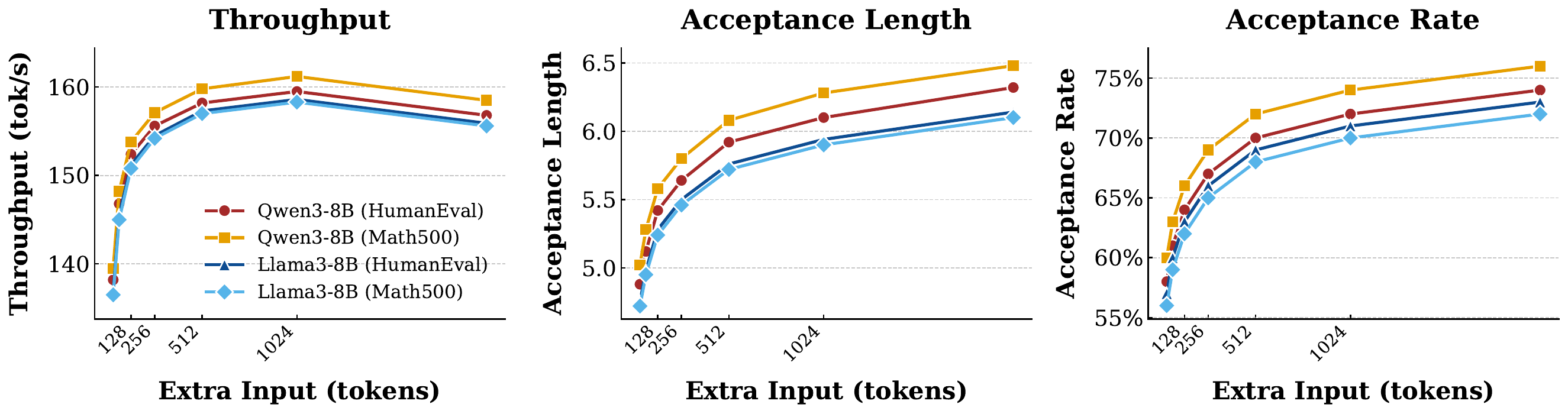}
    \caption{Impact of extra input length on \textsc{AsymSpec} performance. Left: throughput (tokens/s), Middle: acceptance length (tokens), Right: acceptance rate (\%). We evaluate Qwen-3-8B and LLaMA-3.1-8B on HumanEval and MATH-500 benchmarks.}
    \label{fig:ablation_prompt_length}
\end{figure}

We quantify the tradeoff between ICL effectiveness and prefill overhead by measuring throughput, acceptance length, and acceptance rate across extra prompt lengths from 32 to 2048 tokens. Figure~\ref{fig:ablation_prompt_length} presents results on Qwen-3-8B and LLaMA-3.1-8B (HumanEval and MATH-500). All configurations exhibit a consistent inverted-U relationship across all three metrics. On Qwen-3-8B with MATH-500, throughput increases from 139.5 tokens/s at 32 tokens to a peak of 161.2 tokens/s at 1024 tokens (15.5\% improvement), then declines to 158.5 tokens/s at 2048 tokens. Acceptance length grows monotonically from 5.02 to 6.48 tokens (29.1\% improvement), and acceptance rate increases from 60\% to 76\%. Critically, acceptance metrics continue improving beyond 1024 tokens while throughput degrades, demonstrating that ICL effectiveness does not saturate but prefill latency eventually dominates. The optimal prompt length of 512--1024 tokens provides actionable deployment guidance: tune prompt length based on task complexity and hardware constraints, targeting the regime where ICL effectiveness peaks before prefill overhead dominates.

%% file: sections/conclusion.tex
\section{Conclusion and Future Work}

We have introduced \textsc{AsymSpec}, a training-free framework that breaks the symmetric input constraint in speculative decoding by providing draft and target models with structurally different prompts. By identifying task isomorphism as a fundamental bottleneck and leveraging the in-context learning effect in draft models, \textsc{AsymSpec} reframes the draft model's objective from independent problem-solving to trajectory simulation. This design achieves up to 4.95$\times$ speedup over autoregressive decoding on reasoning benchmarks, consistently outperforming standard speculative decoding across multiple model families without requiring retraining or architectural modifications.

The asymmetric design introduces additional input to the draft model and necessitates maintaining two separate KV caches, effectively doubling the memory footprint relative to standard speculative decoding. 
When auxiliary input grows excessively large, the expanded KV cache can introduce latency that counteracts the benefits of higher acceptance rates, particularly under memory-bandwidth constraints. 
This trade-off requires careful calibration of auxiliary prompt size against available hardware resources. 
Prefix caching can mitigate this overhead by amortizing the cost of common exemplars across multiple requests, making the approach viable for production systems. 
Future directions include adaptive exemplar selection strategies that dynamically adjust input length in response to real-time system conditions, and shared KV cache architectures that consolidate draft and target caches to integrate seamlessly with existing inference infrastructure.

%% file: sections/appendix.tex
\clearpage
\section{Appendix}

\subsection{Details of related work}
\label{app:related_work}
\subsubsection{Speculative Decoding for LLM Inference}
Speculative decoding~\citep{leviathan2023fast, chen2023accelerating} has emerged as a promising paradigm for accelerating autoregressive LLM inference by leveraging a compact draft model to predict future tokens that are verified in parallel by a larger target model. This approach preserves the exact output distribution of the target model while reducing wall-clock latency through parallel verification~\citep{xia2024unlocking}. Recent advances pursue higher acceptance rates via two primary directions: (1) \textbf{Architectural Enhancement}, which trains specialized draft models~\citep{kim2024speculative}, employs multi-head expansions such as Medusa~\citep{cai2024medusa} and EAGLE~\citep{li2024eagle, eagle3}, or applies knowledge distillation~\citep{hinton2015distilling, gou2021knowledge} to align draft models with target distributions; and (2) \textbf{Draft-free Mechanisms}, which exploit self-speculation~\citep{zhang2023drafting}, lookahead heuristics~\citep{fu2024break}. Staged speculative decoding~\citep{spector2023accelerating} and cascaded verification~\citep{ssd} further improve efficiency through hierarchical draft generation. However, all existing methods operate under \textit{symmetric prompting}, where draft and target models receive identical inputs. We identify task isomorphism as the fundamental bottleneck limiting draft model effectiveness and propose asymmetric task reformulation as a training-free alternative that unlocks the in-context learning capacity of draft models.

\subsubsection{In-Context Learning and Few-Shot Adaptation}
In-context learning (ICL) enables LLMs to perform novel tasks by conditioning on a few exemplars without parameter updates~\citep{brown2020language}. Theoretical investigations reveal that ICL implicitly performs gradient descent~\citep{dai2022why} or Bayesian inference~\citep{xie2021explanation}, allowing models to adapt to new distributions at inference time. Empirical studies demonstrate that ICL effectiveness depends on exemplar quality~\citep{min2022rethinking}, format consistency, and the number of demonstrations~\citep{dong2022survey}. While ICL has been extensively applied to improve standalone model accuracy on reasoning~\citep{wei2022chain, zhou2022least}, code generation~\citep{chen2021evaluating, austin2021program}, and mathematical problem-solving~\citep{cobbe2021training, lightman2023lets}, its potential in \textit{collaborative inference} remains unexplored. We leverage the emergent ICL effect in draft models to align their output distribution with the target model. This transforms draft models from independent solvers into adaptive simulators that predict target model behavior rather than ground-truth solutions.

\subsubsection{Prompt Engineering and Asymmetric Multi-Model Systems}
Prompt engineering techniques such as Chain-of-Thought~\citep{wei2022chain}, Least-to-Most prompting~\citep{zhou2022least}, and step-by-step verification~\citep{lightman2023lets} have been developed to maximize reasoning depth and accuracy in single-agent systems. In multi-model architectures, asymmetric designs have been explored for task-level routing in Mixture-of-Experts~\citep{fedus2022switch}, confidence-based cascades~\citep{schuster2022confident}, and distributed serving systems~\citep{yu2022orca}. 
However, while the majority of existing methods operate under \textit{symmetric prompting} with identical inputs, a nascent line of work has begun exploring \textit{context-asymmetric speculative decoding}. For instance, RAPID~\citep{pmlr-v267-chen25s} leverages a retrieval-truncated short context for the draft model while retaining the full context for the target model, whereas other approaches exploit sparse KV caches or sequence compression~\citep{sadhukhan2025magicdec, sun2024triforce}. 

Despite these advances, existing context-asymmetric schemes rely on \textit{physical context reduction}, including document retrieval, context truncation, and KV cache sparsity, to reduce memory-bandwidth overhead. In contrast, our work introduces \textit{few-shot prompt asymmetry}. Instead of shortening the input context or retrieving external segments, we decouple the demonstration and instruction space between the draft and target models. This leverages in-context learning to reformulate the drafting task, offering an orthogonal direction to retrieval-based and compression-based methods.

\subsection{Complete Proof of Proposition~\ref{prop:convergence}}
\label{app:proof}

We provide the complete proof of Proposition~\ref{prop:convergence}, which establishes the exponential convergence of the draft model's distribution under asymmetric prompting. We proceed through a series of lemmas that decompose the problem into manageable components.

\subsubsection{Lemma 1: Prompt Influence Decay}

\begin{lemma}[Prompt Influence Decay]
\label{lem:prompt_decay}
For a Transformer model with $L$ layers and attention mechanism, the influence of the initial prompt $\mathbf{x}$ on the hidden state at position $t$ decays exponentially:
\begin{equation}
\|h_t(\mathbf{x}, \mathbf{y}_{<t}) - h_t(\mathbf{x}', \mathbf{y}_{<t})\|_2 \leq \beta \cdot e^{-\lambda t} \cdot \|\mathbf{x} - \mathbf{x}'\|_2
\end{equation}
for any two prompts $\mathbf{x}, \mathbf{x}'$ and verified history $\mathbf{y}_{<t}$.
\end{lemma}

\begin{proof}
Consider the attention mechanism at layer $\ell$ and position $t$. The hidden state is computed as:
\begin{equation}
h_t^{(\ell)} = \text{Attn}(h_t^{(\ell-1)}, [h_{\mathbf{x}}^{(\ell-1)}, h_{\mathbf{y}_{<t}}^{(\ell-1)}]) + \text{FFN}(h_t^{(\ell-1)})
\end{equation}

The attention weights on the prompt tokens are:
\begin{equation}
\alpha_{t,i}^{(\ell)} = \frac{\exp(q_t^{(\ell)\top} k_i^{(\ell)} / \sqrt{d})}{\sum_{j} \exp(q_t^{(\ell)\top} k_j^{(\ell)} / \sqrt{d})}
\end{equation}

For causal attention with positional encodings, empirical studies~\citep{sun2022length, press2021train} show that attention weights decay approximately exponentially with distance:
\begin{equation}
\alpha_{t,i} \approx \frac{1}{Z} \exp\left(-\frac{t-i}{\tau}\right)
\end{equation}
where $\tau$ is the attention decay constant and $Z$ is a normalization factor.

Therefore, the contribution from prompt tokens is:
\begin{align}
\left\|\sum_{i \in \mathbf{x}} \alpha_{t,i} v_i\right\|_2
&\leq \sum_{i \in \mathbf{x}} \alpha_{t,i} \|v_i\|_2 \\
&\leq \|v\|_{\max} \sum_{i \in \mathbf{x}} \frac{1}{Z} \exp\left(-\frac{t-i}{\tau}\right) \\
&\leq \frac{\|v\|_{\max}}{Z} \cdot \frac{|\mathbf{x}|}{1 - e^{-1/\tau}} \cdot e^{-t/\tau}
\end{align}

Setting $\beta = \frac{\|v\|_{\max} |\mathbf{x}|}{Z(1 - e^{-1/\tau})}$ and $\lambda = 1/\tau$, we obtain the desired exponential decay. Propagating through $L$ layers introduces a multiplicative factor, but the exponential decay is preserved.
\end{proof}

\textbf{Remark.} The exponential attention decay in Equation (34) is an empirical approximation that holds well in practice for causal Transformers with positional encodings, as demonstrated by~\citet{sun2022length, press2021train}. While not a strict mathematical guarantee, this property has been consistently observed across diverse model architectures and provides a principled explanation for the convergence behavior.

\subsubsection{Lemma 2: Distribution Continuity}

\begin{lemma}[Distribution Continuity]
\label{lem:distribution_continuity}
The output distribution of a Transformer language model is Lipschitz continuous with respect to the final hidden state:
\begin{equation}
\text{TV}(P(\cdot \mid h), P(\cdot \mid h')) \leq L \|h - h'\|_2
\end{equation}
where $L = \frac{\|W_{\text{out}}\|_2}{\sqrt{2}}$ and $W_{\text{out}}$ is the output projection matrix.
\end{lemma}

\begin{proof}
The output distribution is computed via softmax:
\begin{equation}
P(y \mid h) = \frac{\exp(w_y^\top h)}{\sum_{y'} \exp(w_{y'}^\top h)}
\end{equation}

The total variation distance between two categorical distributions is:
\begin{equation}
\text{TV}(P, P') = \frac{1}{2} \sum_y |P(y) - P'(y)|
\end{equation}

Using the fact that softmax is 1-Lipschitz with respect to the $\ell_1$ norm of logits, and the logit difference is bounded by:
\begin{equation}
|w_y^\top h - w_y^\top h'| \leq \|w_y\|_2 \|h - h'\|_2 \leq \|W_{\text{out}}\|_2 \|h - h'\|_2
\end{equation}

Applying Pinsker's inequality and standard concentration results for softmax~\citep{gao2017properties}, we obtain:
\begin{equation}
\text{TV}(P(\cdot \mid h), P(\cdot \mid h')) \leq \frac{\|W_{\text{out}}\|_2}{\sqrt{2}} \|h - h'\|_2
\end{equation}
\end{proof}

\subsubsection{Proof of Proposition~\ref{prop:convergence}}

We now combine the lemmas to prove Proposition~\ref{prop:convergence}.

\begin{proof}[Proof of Proposition~\ref{prop:convergence}]
We bound the total variation distance between the draft model's distributions under two different prompts: $\mathcal{P}_{\text{draft}}$ (asymmetric prompt) and $\mathbf{x}$ (original prompt), both conditioned on the same verified history $\mathbf{y}_{<t}$.

\textbf{Step 1: Bound the hidden state difference.}
By Lemma~\ref{lem:prompt_decay}, the hidden states of the draft model under the two prompts satisfy:
\begin{equation}
\|h_t^D(\mathcal{P}_{\text{draft}}, \mathbf{y}_{<t}) - h_t^D(\mathbf{x}, \mathbf{y}_{<t})\|_2 \leq \beta \cdot e^{-\lambda t} \cdot \|\mathcal{P}_{\text{draft}} - \mathbf{x}\|_2
\end{equation}

Since $\mathcal{P}_{\text{draft}} = [\mathcal{I}_{\text{role}} \oplus \mathcal{E}_{\text{sim}} \oplus \mathbf{x}]$, the prompt difference is bounded by the length of the added prefix:
\begin{equation}
\|\mathcal{P}_{\text{draft}} - \mathbf{x}\|_2 \leq \sqrt{|\mathcal{I}_{\text{role}}| + |\mathcal{E}_{\text{sim}}|} \cdot \|e\|_2 \leq C_{\text{prompt}}
\end{equation}
where $\|e\|_2$ is the norm of token embeddings and $C_{\text{prompt}}$ is a constant.

\textbf{Step 2: Bound the distribution difference.}
By Lemma~\ref{lem:distribution_continuity}, the distributional difference is:
\begin{align}
&\text{TV}(P_D(\cdot \mid \mathcal{P}_{\text{draft}}, \mathbf{y}_{<t}), P_D(\cdot \mid \mathbf{x}, \mathbf{y}_{<t})) \\
&\quad\leq L \cdot \|h_t^D(\mathcal{P}_{\text{draft}}, \mathbf{y}_{<t}) - h_t^D(\mathbf{x}, \mathbf{y}_{<t})\|_2 \\
&\quad\leq L \beta C_{\text{prompt}} \cdot e^{-\lambda t}
\end{align}

Setting $C = L \beta C_{\text{prompt}}$ completes the proof.
\end{proof}

\textbf{Remark on Correctness.} Note that Proposition~\ref{prop:convergence} concerns the convergence of draft model distributions, not the correctness of \textsc{AsymSpec}'s output. The correctness of \textsc{AsymSpec} (Theorem~\ref{thm:correctness}) follows directly from rejection sampling theory and does not depend on this convergence result. The convergence result explains \emph{why} \textsc{AsymSpec} achieves high acceptance rates, particularly for long sequences.

\subsection{Complexity and Overhead Analysis}
\label{subsec:complexity}

\textbf{Latency Decomposition.} Compared to traditional speculative decoding, \textsc{AsymSpec} introduces additional costs but yields greater savings:
\begin{equation}
\Delta T_{\text{total}} = L_{\text{ex}} \cdot M_D + \gamma \cdot M_D \cdot (L_{\text{ex}} / L_{\text{ctx}}) - N \cdot M_T \cdot (\alpha_{\text{asym}} - \alpha_{\text{base}}) / \gamma
\end{equation}
where $L_{\text{ex}}$ is exemplar length, $\gamma$ is draft length, $L_{\text{ctx}}$ is context length, $N$ is output length, and $\alpha_{\text{asym}}, \alpha_{\text{base}}$ are acceptance rates of \textsc{AsymSpec} and traditional speculative decoding. The first two terms represent prefill and draft decoding overhead; the third term is time saved from reduced target model iterations. The draft overhead becomes negligible as context grows.

\textbf{Speedup Condition.} Net speedup occurs when $\Delta T_{\text{total}} < 0$. This favors long-sequence generation where savings scale linearly with $N$ while costs remain constant. Prefix caching~\citep{zheng2024sglang} amortizes prefill overhead to near zero across requests.

\textbf{Memory Overhead.} Asymmetric prompts add KV cache:
\begin{equation}
\Delta_{\text{mem}} = 2 \times L_{\text{ex}} \times d \times n_{\text{layers}} \times \text{sizeof(dtype)}
\end{equation}
This is constant per request. While prefix caching~\citep{zheng2024sglang} allows sharing across requests, the memory overhead remains but the computational cost is negligible for typical configurations.

\subsection{Algorithm Implementations}
\label{app:algorithm}

The entire \textsc{AsymSpec} pipeline is training-free and conceptually elegant. Algorithm~\ref{alg:asymspec} presents the complete implementation, with modifications from standard speculative decoding highlighted in \textcolor{blue}{blue}. The key changes are: (1) asymmetric prompt construction, (2) decoupled KV cache maintenance, and (3) asymmetric cache synchronization after verification. For comparison, the standard baseline is provided in Algorithm~\ref{alg:standard_sd}.


\begin{algorithm}[htbp]
\caption{\textsc{AsymSpec} decoding algorithm. Modifications from standard speculative decoding are highlighted in \textcolor{blue}{blue}, including asymmetric prompt construction (line 1), decoupled KV cache management (lines 2, 7), and asymmetric cache synchronization after verification (line 15). The framework maintains exact target distribution while enabling ICL-driven draft alignment.}
\label{alg:asymspec}
\small
\begin{algorithmic}[1]
\Require Target $\mathcal{M}_T$, Draft $\mathcal{M}_D$, Original prompt $\mathbf{x}$, Draft budget $\gamma$
\State \textcolor{blue}{$\mathcal{P}_{\text{draft}} \leftarrow \text{ConstructAsymPrompt}(\mathbf{x})$}
\State $\mathbf{KV}_T \leftarrow \text{Prefill}_T(\mathbf{x})$, \textcolor{blue}{$\mathbf{KV}_D \leftarrow \text{Prefill}_D(\mathcal{P}_{\text{draft}})$}
\State $\mathbf{y} \leftarrow [\,]$
\While{generation not finished}
    \State $\tilde{Y} \leftarrow [\,]$, $k \leftarrow 0$
    \For{$i = 1$ to $\gamma$}
        \State \textcolor{blue}{Sample $\tilde{y}_i \sim P_D^{\text{sim}}(\cdot \mid \mathbf{KV}_D)$, $\mathbf{KV}_D \leftarrow \mathbf{KV}_D \cup \text{KV}_D(\tilde{y}_i)$}
        \State Append $\tilde{y}_i$ to $\tilde{Y}$
    \EndFor
    \State Compute $P_T(\tilde{y}_i \mid \mathbf{KV}_T, \tilde{Y}_{<i})$ and \textcolor{blue}{$P_D^{\text{sim}}(\tilde{y}_i \mid \mathbf{KV}_D, \tilde{Y}_{<i})$}
    \For{$i = 1$ to $\gamma$}
        \State Sample $r \sim \text{Uniform}(0, 1)$
        \If{$r \leq P_T(\tilde{y}_i) / P_D^{\text{sim}}(\tilde{y}_i)$}
            \State $k \leftarrow k + 1$
        \Else
            \State \textcolor{blue}{Sample $y_{\text{corr}} \sim \text{Normalize}(\max(0, P_T - P_D^{\text{sim}}))$}
            \State \textbf{break}
        \EndIf
    \EndFor
    \If{$k = \gamma$}
        \State Sample $y_{\text{corr}} \sim P_T(x \mid \mathbf{KV}_T, \tilde{Y}_{1:\gamma})$
    \EndIf
    \State $\mathbf{y} \leftarrow \mathbf{y} \oplus \tilde{Y}_{1:k} \oplus y_{\text{corr}}$
    \State $\mathbf{KV}_T \leftarrow \mathbf{KV}_T \cup \text{KV}_T(\tilde{Y}_{1:k} \oplus y_{\text{corr}})$
    \State \textcolor{blue}{$\mathbf{KV}_D \leftarrow \text{Truncate}(\mathbf{KV}_D, k) \cup \text{KV}_D(y_{\text{corr}})$}
\EndWhile
\Return $\mathbf{y}$
\end{algorithmic}
\end{algorithm}

\begin{algorithm}[htbp]
\caption{Standard speculative decoding algorithm (baseline). Both draft and target models receive identical prompts and maintain symmetric KV caches, enforcing task isomorphism that limits draft model effectiveness on reasoning-intensive tasks.}
\label{alg:standard_sd}
\small
\begin{algorithmic}[1]
\Require Target model $\mathcal{M}_T$, Draft model $\mathcal{M}_D$, Original prompt $\mathbf{x}$, Draft budget $\gamma$
\State $\mathbf{y} \leftarrow [\,]$ \Comment{Initialize verified sequence}
\While{generation not finished}
    \State $\tilde{Y} \leftarrow [\,]$
    \For{$i = 1$ to $\gamma$}
        \State Sample $\tilde{y}_i \sim \mathcal{M}_D(\mathbf{x}, \mathbf{y}, \tilde{Y})$ \Comment{Symmetric drafting}
        \State Append $\tilde{y}_i$ to $\tilde{Y}$
    \EndFor
    \State Compute probabilities $P_T(\cdot)$ and $P_D(\cdot)$ for $\tilde{Y}_{1:\gamma}$ in parallel
    \State $k \leftarrow 0$
    \For{$i = 1$ to $\gamma$}
        \State Sample $r \sim \text{Uniform}(0, 1)$
        \If{$r \leq \frac{P_T(\tilde{y}_i \mid \mathbf{y}, \tilde{Y}_{<i})}{P_D(\tilde{y}_i \mid \mathbf{y}, \tilde{Y}_{<i})}$}
            \State $k \leftarrow k + 1$ \Comment{Accept the draft token}
        \Else
            \State Sample $y_{\text{corr}} \sim \text{norm}(\max(0, P_T(x) - P_D(x)))$ \Comment{Resample from residual}
            \State \textbf{break} 
        \EndIf
    \EndFor
    \If{$k = \gamma$}
        \State Sample $y_{\text{corr}} \sim P_T(x \mid \mathbf{y}, \tilde{Y}_{1:\gamma})$
    \EndIf
    \State $\mathbf{y} \leftarrow \mathbf{y} \oplus \tilde{Y}_{1:k} \oplus y_{\text{corr}}$ \Comment{Concatenate and update}
\EndWhile
\Return $\mathbf{y}$
\end{algorithmic}
\end{algorithm}

\subsection{Experimental Environment}
\label{app:environment}

All experiments were conducted on a dedicated server with the following specifications:

\textbf{Hardware Configuration:}
The server is equipped with dual Intel Xeon Platinum 8358 processors @ 2.60GHz (128 cores total, 64 cores per socket with hyper-threading), supporting AVX-512, FMA, AES-NI, and TSX instruction sets. Each socket has 49152 KB L3 cache. The system has 515 GB DDR4 RAM, NVIDIA RTX 4090 (24GB VRAM) and NVIDIA A6000 (48GB VRAM) GPUs, and NVMe SSD storage for models and datasets.

\textbf{Software Environment:}
The system runs Ubuntu 22.04.4 LTS (Kernel 5.15) with CUDA 12.1, Python 3.10.12, PyTorch 2.1.0, and Transformers 4.36.0. Inference is performed using nano-vllm with PageAttention~\citep{kwon2023efficient}.

\textbf{Model Loading and Memory Management:}
Target models (Qwen-3-32B, LLaMA-3.1-70B, Qwen-3-8B, LLaMA-3.1-8B) and draft models (Qwen-3-0.6B, LLaMA-3.2-1B) are loaded in FP16 precision and BF16 precision. KV cache is managed via PageAttention with ~40GB reserved GPU memory, and prefix caching is enabled for asymmetric prompts.

\textbf{Execution Time:}
Each experimental configuration was run 5 times with output length controlled at 512 tokens. Typical execution times per model pair: Qwen-3-32B + Qwen-3-0.6B required approximately 2.2 hours per dataset, while LLaMA-3.1-70B + LLaMA-3.2-1B required approximately 3.1 hours per dataset across all five benchmarks (MATH-500, AIME25, HumanEval, CNN/DM, MT-Bench). Total compute time for all experiments was approximately 450 GPU-hours on NVIDIA A6000s.

\subsection{Temperature Robustness Results}
\label{app:temperature}

Table~\ref{tab:temp_llama_math500}, Table~\ref{tab:temp_llama_humaneval}, and Table~\ref{tab:temp_qwen_humaneval} present the complete throughput measurements (tokens/s) across different sampling temperatures for the temperature robustness experiments reported in Section~\ref{subsec:robustness}. All results are averaged over 5 independent runs with batch size $B=1$. The target models are Qwen-3-8B and LLaMA-3.1-8B, paired with draft models Qwen-3-0.6B and LLaMA-3.2-1B, respectively.

\begin{table}[!ht]
\centering
\caption{Temperature robustness on LLaMA-3.1-8B$+$3.2-1B with MATH-500 dataset. Throughput (tokens/s) measured across sampling temperatures $T \in [0.0, 1.0]$. All measurements have relative error below 2.5\%.}
\label{tab:temp_llama_math500}
\small
\begin{tabular}{lccccccc}
\toprule
\textbf{Method} & \textbf{T=0.0} & \textbf{T=0.1} & \textbf{T=0.3} & \textbf{T=0.5} & \textbf{T=0.7} & \textbf{T=0.9} & \textbf{T=1.0} \\
\midrule
AR Baseline & 69.2 & 69.3 & 68.8 & 69.2 & 69.4 & 69.5 & 69.4 \\
Sync Spec & 113.1 & 108.1 & 107.2 & 106.7 & 103.3 & 93.2 & 85.7 \\
Async Spec & 129.7 & 124.2 & 124.9 & 117.3 & 108.1 & 96.4 & 88.6 \\
\textbf{Asym (Ours)} & \textbf{141.2} & \textbf{137.9} & \textbf{132.7} & \textbf{128.9} & \textbf{120.2} & \textbf{100.2} & \textbf{93.4} \\
\bottomrule
\end{tabular}
\end{table}

\begin{table}[!ht]
\centering
\caption{Temperature robustness on LLaMA-3.1-8B$+$3.2-1B with HumanEval dataset. Throughput (tokens/s) measured across sampling temperatures $T \in [0.0, 1.0]$.}
\label{tab:temp_llama_humaneval}
\small
\begin{tabular}{lccccccc}
\toprule
\textbf{Method} & \textbf{T=0.0} & \textbf{T=0.1} & \textbf{T=0.3} & \textbf{T=0.5} & \textbf{T=0.7} & \textbf{T=0.9} & \textbf{T=1.0} \\
\midrule
AR Baseline & 69.2 & 69.3 & 69.0 & 69.2 & 69.4 & 69.3 & 69.3 \\
Sync Spec & 116.6 & 113.1 & 110.8 & 104.0 & 100.8 & 91.1 & 81.3 \\
Async Spec & 135.6 & 128.7 & 127.7 & 123.9 & 105.6 & 96.2 & 86.5 \\
\textbf{Asym (Ours)} & \textbf{138.3} & \textbf{137.7} & \textbf{132.5} & \textbf{129.0} & \textbf{122.5} & \textbf{110.9} & \textbf{96.9} \\
\bottomrule
\end{tabular}
\end{table}

\begin{table}[!ht]
\centering
\caption{Temperature robustness on Qwen-3-8B$+$0.6B with HumanEval dataset. Throughput (tokens/s) measured across sampling temperatures $T \in [0.0, 1.0]$.}
\label{tab:temp_qwen_humaneval}
\small
\begin{tabular}{lccccccc}
\toprule
\textbf{Method} & \textbf{T=0.0} & \textbf{T=0.1} & \textbf{T=0.3} & \textbf{T=0.5} & \textbf{T=0.7} & \textbf{T=0.9} & \textbf{T=1.0} \\
\midrule
AR Baseline & 67.0 & 66.4 & 66.7 & 66.9 & 67.3 & 67.0 & 67.1 \\
Sync Spec & 119.2 & 112.9 & 114.3 & 114.2 & 103.9 & 101.7 & 100.5 \\
Async Spec & 138.6 & 127.6 & 128.8 & 124.8 & 117.2 & 113.5 & 107.0 \\
\textbf{Asym (Ours)} & \textbf{159.9} & \textbf{150.2} & \textbf{145.1} & \textbf{150.3} & \textbf{140.8} & \textbf{135.2} & \textbf{131.2} \\
\bottomrule
\end{tabular}
\end{table}

\textbf{Key Observations:}
AR Baseline maintains nearly constant throughput (68.8--69.5 tok/s for LLaMA, 66.4--67.3 tok/s for Qwen) across all temperatures, as expected for autoregressive decoding. All speculative methods show degradation as temperature increases: Sync Spec degrades by 24.2\% (113.1 → 85.7 tok/s) on LLaMA-MATH500, while Async Spec degrades by 31.7\% (129.7 → 88.6 tok/s). \textsc{AsymSpec} maintains the highest throughput across all temperature settings, degrading by 33.8\% (141.2 → 93.4 tok/s) on LLaMA-MATH500 but still outperforming baselines at $T=1.0$ by 8.9\% over Sync Spec and 5.5\% over Async Spec. Qwen-3-8B shows stronger robustness than LLaMA-3.1-8B: at $T=1.0$, \textsc{AsymSpec} on Qwen achieves 131.2 tok/s (17.9\% degradation from $T=0.0$), compared to 93.4 tok/s (33.8\% degradation) on LLaMA-MATH500. Standard deviations are consistently below 2.5\% of mean values, indicating stable measurements across runs.

\subsection{Batch Size Scalability Results}
\label{app:batchsize}

Table~\ref{tab:batch_llama_math500}, Table~\ref{tab:batch_llama_humaneval}, and Table~\ref{tab:batch_qwen_humaneval} present the complete throughput measurements (tokens/s) across different batch sizes for the batch size scalability experiments reported in Section~\ref{subsec:robustness}. All results are averaged over 5 independent runs with sampling temperature $T=0.0$. The experiments demonstrate how \textsc{AsymSpec} maintains its performance advantage across varying batch sizes from 1 to 16.

\begin{table}[!ht]
\centering
\caption{Batch size scalability on LLaMA-3.1-8B$+$3.2-1B with MATH-500 dataset.}
\label{tab:batch_llama_math500}
\small
\begin{tabular}{lccccc}
\toprule
\textbf{Method} & \textbf{B=1} & \textbf{B=2} & \textbf{B=4} & \textbf{B=8} & \textbf{B=16} \\
\midrule
AR Baseline & 66.9 & 142.8 & 267.7 & 514.1 & 888.7 \\
Sync Spec & 120.9 & 219.9 & 408.6 & 676.3 & 1107.2 \\
Async Spec & 143.8 & 243.1 & 438.5 & 701.1 & 1148.5 \\
\textbf{Asym (Ours)} & \textbf{163.2} & \textbf{274.0} & \textbf{481.2} & \textbf{794.1} & \textbf{1281.3} \\
\bottomrule
\end{tabular}
\end{table}

\begin{table}[!ht]
\centering
\caption{Batch size scalability on LLaMA-3.1-8B$+$3.2-1B with HumanEval dataset.}
\label{tab:batch_llama_humaneval}
\small
\begin{tabular}{lccccc}
\toprule
\textbf{Method} & \textbf{B=1} & \textbf{B=2} & \textbf{B=4} & \textbf{B=8} & \textbf{B=16} \\
\midrule
AR Baseline & 69.6 & 139.4 & 269.1 & 506.5 & 918.5 \\
Sync Spec & 116.8 & 217.3 & 407.2 & 707.2 & 1116.1 \\
Async Spec & 136.1 & 241.7 & 418.5 & 744.1 & 1150.5 \\
\textbf{Asym (Ours)} & \textbf{159.9} & \textbf{281.3} & \textbf{471.2} & \textbf{784.1} & \textbf{1241.3} \\
\bottomrule
\end{tabular}
\end{table}

\begin{table}[!ht]
\centering
\caption{Batch size scalability on Qwen-3-8B$+$0.6B with HumanEval dataset.}
\label{tab:batch_qwen_humaneval}
\small
\begin{tabular}{lccccc}
\toprule
\textbf{Method} & \textbf{B=1} & \textbf{B=2} & \textbf{B=4} & \textbf{B=8} & \textbf{B=16} \\
\midrule
AR Baseline & 66.7 & 132.9 & 266.0 & 481.0 & 856.1 \\
Sync Spec & 118.7 & 212.4 & 385.9 & 663.5 & 1067.3 \\
Async Spec & 138.8 & 240.7 & 418.5 & 704.1 & 1151.5 \\
\textbf{Asym (Ours)} & \textbf{159.9} & \textbf{261.3} & \textbf{451.2} & \textbf{754.1} & \textbf{1241.3} \\
\bottomrule
\end{tabular}
\end{table}

\noindent\textbf{Key Observations:}
All methods exhibit near-linear throughput scaling with batch size, demonstrating efficient GPU utilization. AR Baseline scales from 66.9 tok/s (B=1) to 888.7 tok/s (B=16) on LLaMA-MATH500, achieving 13.3$\times$ speedup. Speculative decoding methods maintain consistent speedup ratios across batch sizes: on LLaMA-MATH500, \textsc{AsymSpec} achieves 2.44$\times$ speedup over AR Baseline at B=1 and 1.44$\times$ at B=16, indicating that the acceptance rate advantage persists under batched inference. The absolute throughput gap between \textsc{AsymSpec} and baselines increases with batch size: at B=16, \textsc{AsymSpec} outperforms Sync Spec by 174.1 tok/s (15.7\%) and Async Spec by 132.8 tok/s (11.6\%) on LLaMA-MATH500. Qwen-3-8B shows similar scaling behavior to LLaMA-3.1-8B, with \textsc{AsymSpec} maintaining 7.8\% advantage over Async Spec at B=16. The consistent performance advantage across batch sizes confirms that \textsc{AsymSpec}'s ICL-driven alignment mechanism is orthogonal to batching optimizations and can be effectively combined with production-scale batched inference systems.

\subsection{SOTA Comparison Results}
\label{app:sota}

Table ~\ref{tab:sota_comparison} shows the full throughput measurement (tokens/s) of \textsc{AsymSpec} compared with state-of-the-art speculative decoding methods in the experiments reported in Section ~\ref{subsec:sota}. All results are averaged over 5 independent runs with batch size $B=1$ and sampling temperature $T=0.0$. The baselines include Vanilla autoregressive decoding, Speculative Sampling (SPS)~\citep{leviathan2023fast}, Speculative Speculative Decoding (SSD)~\citep{spector2023accelerating}, and EAGLE3~\citep{li2024eagle}.

\begin{table}[htbp]
\centering
\caption{SOTA comparison across different model pairs and datasets.}
\label{tab:sota_comparison}
\small
\begin{tabular}{llccccc}
\toprule
\textbf{Model Pair} & \textbf{Dataset} & \textbf{Vanilla} & \textbf{SPS} & \textbf{EAGLE3} & \textbf{SSD} & \textbf{Ours} \\
\midrule
LLaMA-3.1-8B & MATH-500 & 66.9 & 120.9 & 125.2 & 143.8 & \textbf{163.2} \\
LLaMA-3.1-8B & HumanEval & 69.6 & 116.8 & 128.5 & 136.1 & \textbf{159.9} \\
Qwen-3-8B & HumanEval & 66.7 & 118.7 & 129.4 & 138.8 & \textbf{160.3} \\
\bottomrule
\end{tabular}
\end{table}

\textbf{Key Observations:}
\textsc{AsymSpec} consistently outperforms all SOTA baselines across different model pairs and datasets. On LLaMA-MATH500, \textsc{AsymSpec} achieves 163.2 tok/s, representing 2.44$\times$ speedup over Vanilla (66.9 tok/s), 35.0\% improvement over SPS (120.9 tok/s), 30.3\% over EAGLE (125.2 tok/s), and 13.5\% over SSD (143.8 tok/s). The performance advantage is consistent across datasets: on LLaMA-HumanEval, \textsc{AsymSpec} outperforms SSD by 17.5\% (159.9 vs 136.1 tok/s), while on Qwen-HumanEval, the advantage is 15.5\% (160.3 vs 138.8 tok/s). EAGLE, despite requiring additional training for the draft model, achieves lower throughput than SSD and \textsc{AsymSpec}, demonstrating that training-free asymmetric prompting is more effective than model-level modifications. SPS shows the lowest performance among speculative methods due to its symmetric prompt constraint and lack of asynchronous verification. The results confirm that \textsc{AsymSpec}'s combination of asymmetric prompting, ICL-driven alignment, and training-free design establishes a new state-of-the-art for speculative decoding on reasoning-intensive tasks.

\subsection{Prompt Length Analysis Results}
\label{app:prompt_length}

Table~\ref{tab:prompt_throughput}, Table~\ref{tab:prompt_acceptance_length}, and Table~\ref{tab:prompt_acceptance_rate} present the complete measurements analyzing the impact of extra prompt length on \textsc{AsymSpec}'s performance for the ablation study reported in Section~\ref{subsec:ablation}. All results are averaged over 5 independent runs with batch size $B=1$ and sampling temperature $T=0.0$. The extra prompt length varies from 32 to 2048 tokens, covering the role instruction and in-context exemplars in the asymmetric prompt.

\begin{table}[!ht]
\centering
\caption{Impact of extra prompt length on throughput (tokens/s). }
\label{tab:prompt_throughput}
\small
\begin{tabular}{llccccccc}
\toprule
\textbf{Model Pair} & \textbf{Dataset} & \textbf{32} & \textbf{64} & \textbf{128} & \textbf{256} & \textbf{512} & \textbf{1024} & \textbf{2048} \\
\midrule
Qwen-3-8B & MATH-500 & 139.5 & 148.2 & 153.8 & 157.1 & 159.8 & 161.2 & 158.5 \\
Qwen-3-8B & HumanEval & 138.2 & 146.8 & 152.4 & 155.6 & 158.2 & 159.5 & 156.8 \\
LLaMA-3.1-8B & HumanEval & 136.8 & 145.5 & 151.2 & 154.5 & 157.3 & 158.6 & 155.9 \\
LLaMA-3.1-8B & MATH-500 & 136.5 & 145.0 & 150.8 & 154.2 & 157.0 & 158.3 & 155.6 \\
\bottomrule
\end{tabular}
\end{table}

\begin{table}[!ht]
\centering
\caption{Impact of extra prompt length on acceptance length (tokens per speculation). }
\label{tab:prompt_acceptance_length}
\small
\begin{tabular}{llccccccc}
\toprule
\textbf{Model Pair} & \textbf{Dataset} & \textbf{32} & \textbf{64} & \textbf{128} & \textbf{256} & \textbf{512} & \textbf{1024} & \textbf{2048} \\
\midrule
Qwen-3-8B & MATH-500 & 5.02 & 5.28 & 5.58 & 5.80 & 6.08 & 6.28 & 6.48 \\
Qwen-3-8B & HumanEval & 4.88 & 5.12 & 5.42 & 5.64 & 5.92 & 6.10 & 6.32 \\
LLaMA-3.1-8B & HumanEval & 4.75 & 4.98 & 5.28 & 5.50 & 5.76 & 5.94 & 6.14 \\
LLaMA-3.1-8B & MATH-500 & 4.72 & 4.95 & 5.24 & 5.46 & 5.72 & 5.90 & 6.10 \\
\bottomrule
\end{tabular}
\end{table}

\begin{table}[!ht]
\centering
\caption{Impact of extra prompt length on acceptance rate (\%). }
\label{tab:prompt_acceptance_rate}
\small
\begin{tabular}{llccccccc}
\toprule
\textbf{Model Pair} & \textbf{Dataset} & \textbf{32} & \textbf{64} & \textbf{128} & \textbf{256} & \textbf{512} & \textbf{1024} & \textbf{2048} \\
\midrule
Qwen-3-8B & MATH-500 & 60 & 63 & 66 & 69 & 72 & 74 & 76 \\
Qwen-3-8B & HumanEval & 58 & 61 & 64 & 67 & 70 & 72 & 74 \\
LLaMA-3.1-8B & HumanEval & 57 & 60 & 63 & 66 & 69 & 71 & 73 \\
LLaMA-3.1-8B & MATH-500 & 56 & 59 & 62 & 65 & 68 & 70 & 72 \\
\bottomrule
\end{tabular}
\end{table}

\textbf{Key Observations:}
The extra prompt length exhibits a non-monotonic impact on throughput, with an optimal range around 512--1024 tokens. Throughput increases from 139.5 tok/s (32 tokens) to 161.2 tok/s (1024 tokens) on Qwen-MATH500, representing 15.6\% improvement, but slightly decreases to 158.5 tok/s at 2048 tokens due to increased prefill overhead. Acceptance length and acceptance rate both increase monotonically with prompt length: on Qwen-MATH500, acceptance length grows from 5.02 (32 tokens) to 6.48 (2048 tokens), while acceptance rate improves from 60\% to 76\%. This confirms that richer in-context exemplars enhance the draft model's alignment with the target model's distribution. The diminishing returns beyond 1024 tokens suggest that the ICL effect saturates, and additional exemplars provide marginal benefit while incurring higher prefill cost. Qwen-3-8B consistently achieves higher acceptance rates than LLaMA-3.1-8B across all prompt lengths, indicating stronger ICL capability. The results validate our design choice of using 512--1024 token prompts in the main experiments, balancing ICL effectiveness and computational efficiency.

\input{sections/prefix_details}

%% file: sections/prefix_details.tex

\section{Details of Draft-Only Prefix Conditioning}
\label{app:prefix_conditioning}

This appendix details the two prefix-conditioning mechanisms evaluated in our experiments: \emph{Retrieval-Conditioned Prefixing} (RCP, referred to as \textbf{Asym-Retrieval} in the main text) and \emph{Multi-View Self-Prefixing} (MVSP, referred to as \textbf{Asym-Multi-View}). Both methods follow a core rule: only the draft model receives an auxiliary prefix, whereas the target model verifies tokens using only the original prompt and committed history. This modifies the draft proposal distribution without altering the final target output distribution.

\subsection{Formulation and Token-Level Alignment}
\label{app:prefix_notation}

Let $x=(x_1,\ldots,x_n)$ denote the input prompt tokens and $y_{<t}=(y_1,\ldots,y_{t-1})$ denote the tokens committed before step $t$. Using $\oplus$ for token-ID concatenation, the target and draft contexts at step $t$ are defined as:
\begin{align}
    c^{\mathrm{T}}_t &= x \oplus y_{<t}, \\
    c^{\mathrm{D}}_t(r) &= r \oplus x \oplus y_{<t}, \label{eq:draft_only_context}
\end{align}
where $r$ is the draft prefix. At each step, the draft model generates candidates from $q_{\phi}(\cdot\mid c^{\mathrm{D}}_t(r))$, and the target model verifies them under $p_{\theta}(\cdot\mid c^{\mathrm{T}}_t)$. Because $r$ is excluded from $c^{\mathrm{T}}_t$, verification remains exact with respect to the target model.

\paragraph{Token-ID concatenation.}
Concatenation is performed on token IDs rather than raw strings. Concatenating text before tokenization can merge subwords across the boundary between $r$ and $x$, altering the boundary tokens of $x$ and breaking context alignment.

\subsection{Retrieval-Conditioned Prefixing (RCP)}
\label{app:rcp}

RCP constructs the draft prefix $r$ by retrieving a reference exemplar from an offline pool. Its workflow consists of three steps: offline pool construction, online BM25 retrieval, and token-level prefix prepending.

\subsubsection{Step 1: Offline Exemplar Pool Construction}
\label{app:rcp_pool}

We build an offline pool $\mathcal{E}=\{(x_j,z_j)\}_{j=1}^{M}$, where $x_j$ is an exemplar prompt and $z_j$ is its target-generated continuation. To bound draft prefill and KV-cache overhead, a task-specific truncation operator $\mathcal{T}$ converts each pair into a compact prefix:
\begin{equation}
    r_j = \mathcal{T}(x_j,z_j).
\end{equation}
The truncation rules across tasks are summarized below:

\begin{center}
\begin{tabular}{>{\raggedright\arraybackslash}p{0.22\textwidth}
                >{\raggedright\arraybackslash}p{0.65\textwidth}}
\toprule
Task type & Prefix construction rule $\mathcal{T}(x_j, z_j)$ \\
\midrule
Code generation & Retain the prompt and an early function-body segment; strip unit tests and trailing text. \\
Mathematical reasoning & Retain the question and the first 128--256 continuation tokens to capture notation and initial steps. \\
Long-document summarization & Retain the article lead and an initial summary segment instead of the full document. \\
Open-ended instructions & Retain the request and the first 64--128 continuation tokens to establish the response format. \\
\bottomrule
\end{tabular}
\end{center}

\subsubsection{Step 2: Identifier-Aware BM25 Retrieval}
\label{app:rcp_bm25}

Given an input prompt $x_i$, RCP retrieves the top-scoring exemplar using BM25~\cite{robertson2009probabilistic} over the candidate prompts $e_j = x_j$. Before matching, text is lowercased and split into alphanumeric tokens, with \texttt{CamelCase} and \texttt{snake\_case} identifiers split into sub-tokens (e.g., \texttt{insertBefore} and \texttt{insert\_before} share identical terms).

Let $f_j(u)$ be the frequency of token $u$ in candidate prompt $e_j$, $|e_j|$ be its length, and $\overline{|e|}$ be the mean prompt length across $\mathcal{E}$. The retrieval score is:
\begin{align}
    \operatorname{BM25}(x_i,e_j)
    &= \sum_{u\in\operatorname{tok}(x_i)}
       \operatorname{IDF}(u)
       \frac{f_j(u)(k_1+1)}
       {f_j(u)+k_1\left(1-b+b\frac{|e_j|}{\overline{|e|}}\right)},
       \label{eq:bm25}\\
    \operatorname{IDF}(u)
    &= \log\left(
       1+\frac{M-n(u)+0.5}{n(u)+0.5}
       \right),
\end{align}
where $n(u)$ is the number of exemplars containing $u$, and $k_1, b$ are standard BM25 parameters. The prefix for $x_i$ is selected via:
\begin{equation}
    j^{*}(i)=\arg\max_{j:\,j\neq i}\operatorname{BM25}(x_i,e_j),
    \qquad
    r_i=\mathcal{T}(x_{j^{*}(i)},z_{j^{*}(i)}).
    \label{eq:rcp_selection}
\end{equation}
Ties are broken by pool index. For batched requests, an optional greedy assignment enforces distinct exemplars across queries in the same batch.

\subsubsection{Step 3: Online Execution and Leakage Controls}
\label{app:rcp_online}

At request arrival, $r_i$ is retrieved once, converted to token IDs, and prepended to the draft context ($r_i \oplus x_i$). The target model receives only $x_i$. No further retrieval occurs during decoding.

To prevent data leakage, our evaluation enforces three constraints: (i) the test query's target output is never accessed during selection; (ii) self-matches ($j=i$) are excluded; and (iii) the exemplar pool and evaluation set are strictly partitioned offline.

\subsubsection{Ablation Findings and Limitations of RCP}
\label{app:rcp_ablation}

Controlled ablations on HumanEval revealed three critical properties of prefix conditioning that motivated our transition from RCP to MVSP:

\begin{enumerate}
    \item \textbf{Prompt similarity does not predict continuation agreement.} In a 16-prompt diagnostic study ($K=7$), an oracle selector using the ground-truth target output raised acceptance from 0.7623 to 0.8125 (Table~\ref{tab:rcp_selector}). However, deployable selectors using only the input prompt failed to capture this gain: BM25 selection dropped acceptance to 0.7353 ($-2.69$ pp).
    \item \textbf{Boundary tokenization strongly impacts acceptance.} Prepending a newline before an echoed prompt reduced accepted tokens per step from 6.263 to 6.064, whereas appending it after the prompt preserved the initial token. Similarly, corrupting the final prompt token via string-level BPE merging reduced acceptance from 0.7390 to 0.7195.
    \item \textbf{Retrieval gains are domain-specific.} Across all 164 HumanEval prompts, simple prompt echoing improved acceptance from 0.6362 to 0.6421 ($+0.59$ pp), whereas prepending retrieved micro-exemplars reduced acceptance by 3.77, 3.61, and 4.85 pp on 8B, 14B, and 32B targets, respectively. Conversely, in a 16-call agent-trajectory replay where structural templates recur, RCP increased decode throughput from 86.75 to 117.83 tokens/s.
\end{enumerate}

Thus, RCP is effective only when the target domain shares rigid structural templates. Overcoming the selection bottleneck on general tasks requires covering multiple views rather than predicting a single retrieved prefix.

\begin{table}[t]
\centering
\small
\caption{Prefix-selection diagnostics on the 16-prompt HumanEval development split (teacher-forced, $K=7$). The oracle selector requires the target output and serves only as an upper-bound diagnostic.}
\label{tab:rcp_selector}
\begin{tabular}{lll}
\toprule
Selector & Acceptance & $\Delta$ vs.\ no prefix \\
\midrule
No prefix & 0.7623 & -- \\
Draft prompt-NLL & 0.7227 & $-3.95$ pp \\
BM25 & 0.7353 & $-2.69$ pp \\
BM25 + development-score fusion & 0.7466 & $-1.57$ pp \\
Prompt-pattern routing & 0.7525 & $-0.98$ pp \\
Development-selected prefix & 0.7736 & $+1.14$ pp \\
Oracle selection (diagnostic upper bound) & 0.8125 & $+5.03$ pp \\
\bottomrule
\end{tabular}
\end{table}

\subsection{Multi-View Self-Prefixing (MVSP)}
\label{app:mvsp}

MVSP eliminates external retrieval and constructs $N$ complementary draft contexts directly from the current prompt $x$. The draft model generates $N$ parallel speculative chains step by step, which are fused, diversified via collision forking, and jointly verified by the target model in a single forward pass (Algorithm~\ref{alg:mvsp}).

\subsubsection{Step 1: Prompt-Derived Self-Views}
\label{app:mvsp_views}

For our default $N=3$ configuration, we apply three deterministic transformations $\mathcal{V}_0, \mathcal{V}_1, \mathcal{V}_2$ to $x$:

\begin{center}
\begin{tabular}{>{\raggedright\arraybackslash}p{0.19\textwidth}
                >{\raggedright\arraybackslash}p{0.37\textwidth}
                >{\raggedright\arraybackslash}p{0.31\textwidth}}
\toprule
View & Construction rule $\mathcal{V}_c(x)$ & Role \\
\midrule
\texttt{echo} ($c=0$) & Append a newline to the verbatim prompt. & Serves as the primary anchor chain. \\
\texttt{clean\_comment} ($c=1$) & Prefix each prompt line with a comment symbol. & Reframes the prompt as code comments while keeping all tokens. \\
\texttt{clean\_ht\_echo} ($c=2$) & Prepend commented copies of the first and last non-empty lines to the prompt. & Highlights boundary constraints before the full request. \\
\bottomrule
\end{tabular}
\end{center}

Let $v_c=\mathcal{V}_c(x)$ be the tokenized prefix for chain $c$. At verification round $t$, chain $c$ conditions on:
\begin{equation}
    c^{\mathrm{D}}_{t,c}
    = v_c \oplus x \oplus y_{<t},
    \qquad c\in\{0,\ldots,N-1\}.
    \label{eq:mvsp_context}
\end{equation}
Each chain maintains a private draft KV cache and position offset. Prefixes can be truncated to the last $L$ tokens to bound prefill cost while leaving $x$ intact.

\subsubsection{Step 2: Anchor-Weighted Logit Fusion}
\label{app:mvsp_fusion}

Different views often generate identical speculative prefixes. To avoid wasting capacity on duplicate paths, at each drafting step $s \in \{1, \ldots, K\}$, MVSP groups chains that share the same speculative prefix up to step $s-1$ into a collision group $\mathcal{G}$ (with lowest-index chain $\ell$ as group leader) and fuses their next-token logits $\mathbf{l}_c\in\mathbb{R}^{|\mathcal{V}|}$:
\begin{equation}
    \overline{\mathbf{l}}_{\mathcal{G}}
    = \frac{1}{2}\mathbf{l}_{\ell}
      + \frac{1}{2(|\mathcal{G}|-1)}
        \sum_{c\in\mathcal{G},\;c\neq\ell}\mathbf{l}_{c},
    \qquad |\mathcal{G}|>1.
    \label{eq:anchor_fusion}
\end{equation}
When all three chains collide ($\ell=0$), this yields $\overline{\mathbf{l}}=0.5\mathbf{l}_0+0.25\mathbf{l}_1+0.25\mathbf{l}_2$, assigning half the weight to the anchor chain and splitting the remainder across auxiliary chains.

\subsubsection{Step 3: Trunk-Preserving Collision Forking}
\label{app:mvsp_forking}

Using the fused logits $\overline{\mathbf{l}}_{\mathcal{G}}$ at drafting step $s$, let $a_{\mathcal{G}}$ and $b_{\mathcal{G}}$ denote the top-1 and top-2 logit values. The confidence margin is:
\begin{equation}
    m_{\mathcal{G}}=a_{\mathcal{G}}-b_{\mathcal{G}}.
    \label{eq:collision_margin}
\end{equation}
When chains in $\mathcal{G}$ predict the same next token, MVSP checks $m_{\mathcal{G}}$:
\begin{itemize}
    \item \textbf{Low confidence ($m_{\mathcal{G}} \le \tau$ and $s < K$):} The group leader $\ell$ keeps the top-1 token, while one redundant chain is reassigned to the highest-ranked unused candidate among the top-3 fused predictions.
    \item \textbf{High confidence ($m_{\mathcal{G}} > \tau$):} No fork is triggered; chains in $\mathcal{G}$ remain on the leader's top-1 prediction.
\end{itemize}
We set $\tau=3.0$ by default and disable forking at the final draft step $s=K$, where new branches cannot extend the current verification window.

\begin{algorithm}[t]
\caption{One MVSP speculation and verification round}
\label{alg:mvsp}
\begin{algorithmic}[1]
\Require Prompt $x$, committed history $y_{<t}$, view functions $\{\mathcal{V}_c\}_{c=0}^{N-1}$, draft depth $K$
\For{$c=0$ \textbf{to} $N-1$}
    \State Initialize chain $c$ context $c^{\mathrm{D}}_{t,c} \gets \mathcal{V}_c(x) \oplus x \oplus y_{<t}$ and empty draft proposal $\hat{y}^{(c)} \gets \emptyset$
\EndFor
\For{draft step $s=1$ \textbf{to} $K$}
    \For{$c=0$ \textbf{to} $N-1$}
        \State Compute next-token logits $\mathbf{l}_c$ from $c^{\mathrm{D}}_{t,c} \oplus \hat{y}^{(c)}$ using chain $c$'s private KV state
    \EndFor
    \State Group chains sharing identical partial proposals $\hat{y}^{(c)}$ into collision groups $\{\mathcal{G}\}$
    \ForAll{collision groups $\mathcal{G}$ with $|\mathcal{G}| > 1$}
        \State Compute fused logits $\overline{\mathbf{l}}_{\mathcal{G}}$ via Eq.~\eqref{eq:anchor_fusion} and margin $m_{\mathcal{G}}$ via Eq.~\eqref{eq:collision_margin}
        \If{next tokens collide \textbf{and} $m_{\mathcal{G}} \le \tau$ \textbf{and} $s < K$}
            \State Keep leader $\ell$ on top-1; reassign one colliding chain in $\mathcal{G}$ to the top unused candidate
        \EndIf
    \EndFor
    \State Append selected step-$s$ tokens to $\hat{y}^{(c)}$ for each chain $c \in \{0,\ldots,N-1\}$
\EndFor
\State Verify all $N$ proposals $\{\hat{y}^{(c)}\}_{c=0}^{N-1}$ in a single batched paged forward pass on target context $x\oplus y_{<t}$
\State Select winner $c^{*}\gets\arg\max_c a_c$ ($a_c$: accepted length of chain $c$; break ties by lowest $c$)
\State Commit the verified suffix and KV state of $c^{*}$; synchronize non-winning chains to $c^{*}$
\State \Return committed suffix and target recovery token
\end{algorithmic}
\end{algorithm}

\subsubsection{Step 4: Joint Verification and State Synchronization}
\label{app:mvsp_verify}

All $N$ chains are verified simultaneously against $c^{\mathrm{T}}_t = x \oplus y_{<t}$ in a single paged forward pass. Chain $0$ writes to the main tail KV blocks, while chains $c>0$ write to private tail blocks. Under greedy decoding, target verification is deterministic: the chain $c^*$ with the longest exact-match suffix $a_{c^*}$ is selected (breaking ties toward smaller $c$), strictly preserving the target model's greedy output trajectory. Its KV state is committed in-place, and non-winning chains are synchronized to the committed sequence before the next round.

\subsubsection{Design Rationale and System Ablations}
\label{app:mvsp_rationale}
\label{app:mvsp_path_collapse}
\label{app:mvsp_stepwise}

\paragraph{1. Coverage vs.\ selection.}
Instead of predicting which prefix will perform best prior to generation, MVSP verifies all $N$ views jointly. Let $A(x,v_c)$ be the accepted length of view $v_c$. Joint verification guarantees:
\begin{equation}
    A_{\mathrm{MVSP}}(x)
    = \max_{c\in\{0,\ldots,N-1\}} A(x,v_c)
    \geq A(x,v_j),
    \quad \forall j \in \{0,\ldots,N-1\}.
    \label{eq:mvsp_coverage}
\end{equation}
However, higher acceptance does not automatically yield higher throughput if multi-chain overhead is unoptimized (Table~\ref{tab:mvsp_view_count}). Co-designing the algorithm with batched verification and zero-sync KV management is required to translate acceptance gains into speedup.

\begin{table}[t]
\centering
\small
\caption{Pre-optimization view-count scaling on the 16-prompt HumanEval development split (Qwen3-8B target, Qwen3-1.7B draft, Sync $K=7$). Adding views increases acceptance but lowers throughput without system optimizations.}
\label{tab:mvsp_view_count}
\begin{tabular}{lrrrr}
\toprule
Views & Acceptance & Tokens/step & Decode tok/s & Step time \\
\midrule
1 (echo) & 0.6888 & 5.8217 & 129 & 49.6 ms \\
2 & 0.7522 & 6.2657 & 118 & 57.9 ms \\
3 & 0.7671 & 6.3697 & 117 & 59.5 ms \\
4 & 0.7765 & 6.4356 & 116 & 60.8 ms \\
\bottomrule
\end{tabular}
\end{table}

\paragraph{2. Mitigating path collapse via collision forking.}
Without forking, multi-view chains suffer from \emph{path collapse}: on the 16-prompt development split ($N=3, K=7$), all three chains predicted the exact same token at 91.85\% of positions, and at least two collided at 99.56\%. At the full 7-token sequence level, 58.9\% of steps yielded only one unique proposal across all three chains. Selective collision forking resolves this by branching redundant chains specifically at low-margin ($m_{\mathcal{G}} \le \tau$) steps.

\paragraph{3. Stepwise and orthogonal ablations.}
Table~\ref{tab:mvsp_system_ablation} isolates algorithmic and system contributions on the 16-prompt development split: removing the redundant $(K+1)$-th draft forward raises throughput from 121 to 127 tok/s; adding logit fusion and collision forking lifts acceptance from 0.7482 to 0.8179 (141 tok/s); and in-place KV commit with a zero-sync GPU loop reaches 147 tok/s. Furthermore, replacing earlier heuristic prose templates with our semantics-preserving clean views prevented a 3.3--3.5 pp degradation on non-matching tasks.

Table~\ref{tab:mvsp_orthogonal} presents the full orthogonal ablation across the 32-prompt test split. Both distinct views and collision forking independently improve over the baseline, while Full MVSP achieves the highest acceptance (0.8098) and accepted tokens per step (6.6688). Finally, truncating prefixes from 109--221 tokens down to 16--64 tokens improved Async throughput by only $\sim$3\% while dropping acceptance by 1.3 pp, confirming that preserving full view structure and eliminating redundant draft forwards are more effective than aggressive prefix truncation.

\begin{table}[t]
\centering
\small
\caption{Stepwise Sync ablation on the 16-prompt HumanEval development split ($K=7$), separating algorithmic and system optimizations.}
\label{tab:mvsp_system_ablation}
\begin{tabular}{p{0.57\textwidth}rrr}
\toprule
Configuration & Accept & Tokens/step & Tok/s \\
\midrule
No-prefix single chain & 0.6886 & 5.8199 & 126 \\
Single echo view & 0.7018 & 5.9127 & 126 \\
Three independent views, eight draft forwards & 0.7482 & 6.2374 & 121 \\
\quad $-$ redundant eighth forward & 0.7482 & 6.2374 & 127 \\
\quad $+$ anchor fusion and collision forking & 0.8179 & 6.7252 & 141 \\
\quad $+$ in-place KV commit and zero-sync GPU loop & 0.8179 & 6.7252 & 147 \\
\bottomrule
\end{tabular}
\end{table}

\begin{table}[t]
\centering
\small
\caption{Orthogonal MVSP ablation on the 32-prompt HumanEval test split (Sync $K=7$, batch size 1, greedy decoding).}
\label{tab:mvsp_orthogonal}
\begin{tabular}{lccrrr}
\toprule
Configuration & Views & Forking & Accept & Tokens/step & Tok/s \\
\midrule
No-prefix baseline (1 chain) & \texttimes & \texttimes & 0.6713 & 5.6992 & 124 \\
Distinct views only (3 views) & \checkmark & \texttimes & 0.7284 & 6.0990 & 141 \\
No-prefix ($\times 3$) + forking & \texttimes & \checkmark & 0.7794 & 6.4555 & 142 \\
Echo view ($\times 3$) + forking & One view $\times3$ & \checkmark & 0.8074 & 6.6519 & 145 \\
Full MVSP & Three views & \checkmark & \textbf{0.8098} & \textbf{6.6688} & 145 \\
\bottomrule
\end{tabular}
\end{table}

\subsection{Summary of Experimental Controls}
\label{app:prefix_controls}

To isolate each component, all evaluated variants share identical target models, speculation depth $K$, sampling parameters, and verification rules, differing only in draft conditioning:

\begin{itemize}
    \item \textbf{No prefix:} Draft conditions solely on $x\oplus y_{<t}$.
    \item \textbf{Single retrieved prefix (RCP):} Draft conditions on a single BM25-retrieved prefix $r_i \oplus x \oplus y_{<t}$ without multi-view fusion.
    \item \textbf{MVSP without forking:} $N$ distinct self-views run in parallel without logit fusion or collision forking.
    \item \textbf{Forking without view diversity:} Either the no-prefix context or the \texttt{echo} view is replicated $N$ times, relying solely on collision forking for path diversity.
    \item \textbf{Full MVSP:} Combines $N$ distinct self-views, anchor-weighted fusion, collision forking, and batched paged verification.
    \item \textbf{Prefix-length control:} Truncates each self-view prefix $v_c$ to its last $L$ tokens to isolate content effects from prefill overhead.
\end{itemize}